\documentclass[numberedappendix, iop, apj, revtex4]{emulateapj}
\usepackage{natbib}
\usepackage{url}
\usepackage{amsmath} 
\usepackage{appendix}
\usepackage{hyperref} 

\def\ciis{C~II$^{*}$}
\def\ciisl{C~II$^{*}$$\lambda$1335.7}
\def\lc{${\ell_{c}}$}

\def\smpykpc{$M_{\odot}{\rm \ yr^{-1} \ kpc^{-2}}$}

\begin{document}

\shorttitle{SFR Efficiency in H\textsc{i} gas from $z\sim1$ to $z\sim3$}
\shortauthors{Rafelski, Gardner, Fumagalli, Neeleman, Teplitz, et al.}

\title{The Star Formation Rate Efficiency of Neutral Atomic-dominated Hydrogen Gas in the Outskirts of Star Forming Galaxies from $z\sim1$ to $z\sim3$}

\author{
  Marc Rafelski\altaffilmark{1,2},
  Jonathan P. Gardner\altaffilmark{1},
  Michele Fumagalli\altaffilmark{3},
  Marcel Neeleman\altaffilmark{4}, 
  Harry I. Teplitz\altaffilmark{5}, 
  Norman Grogin\altaffilmark{6},
  Anton M. Koekemoer\altaffilmark{6}, 
  Claudia Scarlata\altaffilmark{7}
}

\altaffiltext{1} {Goddard Space Flight Center, Code 665, Greenbelt, MD 20771, USA marc.a.rafelski@nasa.gov}
\altaffiltext{2}{NASA Postdoctoral Program Fellow}
\altaffiltext{3}{Institute for Computational Cosmology and Centre for Extragalactic Astronomy, Department of Physics, Durham University, South Road, Durham DH1 3LE, UK}
\altaffiltext{4}{Department of Astronomy \& Astrophysics, UCO/Lick Ob- servatory, 1156 High Street, University of California, Santa Cruz, CA 95064, USA}
\altaffiltext{5}{Infrared Processing and Analysis Center, MS 100-22, Caltech, Pasadena, CA 91125, USA}
\altaffiltext{6}{Space Telescope Science Institute, 3700 San Martin Drive Baltimore, MD 21218}
\altaffiltext{7}{Minnesota Institute for Astrophysics, School of Physics and Astronomy, University of Minnesota, Minneapolis, MN 55455}

\begin{abstract}

Current observational evidence suggests that the star formation rate (SFR) efficiency of neutral atomic hydrogen gas measured in Damped Ly$\alpha$ Systems (DLAs) at $z\sim$3 is more than 10 times lower than predicted by the Kennicutt-Schmidt (KS) relation. To understand the origin of this deficit, and to investigate possible evolution with redshift and galaxy properties, we measure the SFR efficiency of atomic gas at $z\sim$1, $z\sim$2, and $z\sim$3 around star-forming galaxies. We use new robust photometric redshifts in the Hubble Ultra Deep Field to create galaxy stacks in these three redshift bins, and measure the SFR efficiency by combining DLA absorber statistics with the observed rest-frame UV emission in the galaxies' outskirts. We find that the SFR efficiency of H\textsc{i} gas at $z>1$ is $\sim$1-3\% of that predicted by the KS relation. Contrary to simulations and models that predict a reduced SFR efficiency with decreasing metallicity and thus with increasing redshift, we find no significant evolution in the SFR efficiency with redshift. Our analysis instead suggests that the reduced SFR efficiency is driven by the low molecular content of this atomic-dominated phase, with metallicity playing a secondary effect in regulating the conversion between atomic and molecular gas. This interpretation is supported by the similarity between the observed SFR efficiency and that observed in local atomic-dominated gas, such as in the outskirts of local spiral galaxies and local dwarf galaxies.

\end{abstract}


\pagebreak
\clearpage

\section{Introduction}
\label{intro}

The cold gas from which stars form is difficult to detect in emission at high redshift, and therefore 
is generally studied in absorption to background quasars (QSOs), particularly with Damped Ly$\alpha$ systems (DLAs), which have a minimum H\textsc{i} column density of $2\times10^{20}$ cm$^{-2}$. DLAs are found to
trace most of the neutral H\textsc{i} gas in the Universe, containing enough gas to account for 50\% 
of the mass content of visible matter in modern galaxies \citep{Wolfe:2005}.
In addition to studying the individual physical properties of absorbers, surveys of DLAs at high redshift have found that the comoving average H\textsc{i} mass density decreases by a factor of 2 from 
$z\sim5$ to $z\sim2$ \citep{Prochaska:2009, Noterdaeme:2012, Crighton:2015, SanchezRamirez:2015} and
the cosmic metallicity of the H\textsc{i} gas is observed to increase by a factor of 8 from 
$z\sim5$ to $z\sim1$ \citep{Prochaska:2003a, Rafelski:2012, Rafelski:2014}.
Yet even with these advances we know little about the extent and size of the absorbing gas.

Simulations suggest that DLAs trace dense gas in and around galaxies \citep[e.g.][]{Nagamine:2007, Pontzen:2008, Tescari:2009, Cen:2012, Bird:2014}.
Measurements of the cross-correlation of the Ly$\alpha$ forest and DLAs find a large bias factor which implies host halo masses of DLAs as high as $\sim10^{12}M_\odot$ at $z>2.15$ \citep{FontRibera:2012}. 
At the same time, the cross-correlation of DLAs and Lyman break galaxies at $z\sim3$ suggest host halo masses of $\sim10^{11}M_\odot$ \citep{Cooke:2006}.
At $z\sim3$, star forming galaxies (SFGs) have halo masses of $\sim10^{11.5}M_\odot$ \citep{Bielby:2013}, 
indicating that DLAs could be associated with
typical or even massive SFGs. 

Direct imaging of high redshift DLAs has primarily failed due to a combination of intrinsically low 
star formation rates (SFRs) of the associated galaxies and the glare of the background QSO. The bright QSOs prevent 
searches for faint galaxy associations at $\lesssim$1 arcsec from the absorbing gas \citep[e.g.][]{Warren:2001}. 
Therefore, despite almost 20 years of observations, only about a dozen DLA host galaxies have been observed in 
emission (see tables 1 \& 2 in \citet{Krogager:2012} and \citet{Fumagalli:2015a} respectively). 
Although these galaxies appear to be mostly typical SFGs \citep{Moller:2002a, Fynbo:2010,Fynbo:2013, Peroux:2012, Bouche:2013, Krogager:2013, Jorgenson:2014b},  a publication bias exists due to a lack of census for non-detections and biased search criteria (e.g. targeting high metallicity systems, and with sensitivities only sufficient to observe more luminous galaxies). Recently,  \citet{Fumagalli:2015a} provided a full census not biased by metallicity and did not detect the host galaxies. 

There are currently four other techniques available to investigate the in-situ SFR of DLAs: 
1) gamma ray burst (GRB) observations, 2) the `double' DLA technique, 3) the CII* technique, and 4) a statistical comparison of absorbers with observed emission.
In the next few paragraphs we will will briefly describe these techniques.

The use of GRBs instead of QSOs to study DLAs is promising, as GRBs fade over time, enabling deep searches for emission from the
DLAs without the bright nearby quasar. 
The majority of GRB-DLAs are not intervening, but rather are associated with the galaxies hosting the GRB. Therefore, 
GRB-DLAs will be biased with respect to their selection, as GRBs are related to supernova explosions within galaxies \citep[e.g.][]{Kumar:2015}, and therefore are {\it a priori} associated with SFGs. 
GRB-DLAs have slightly different physical properties than QSO-DLAs, such as a different metallicity evolution \citep{Cucchiara:2015},
and only a small number of GRB-DLA sightlines have been studied, since the afterglows fade over time \citep{Schulze:2012}. 

The double DLA technique uses QSO sightlines containing two optically-thick absorbers, with the higher redshift system acting as 
a blocking filter for the lower redshift source. One can then conduct a sensitive search for emission from the lower redshift source, as the higher redshift system blocks all light from the QSO \citep{Steidel:1992, OMeara:2006,Fumagalli:2010a, Fumagalli:2014b}.
An unbiased search for DLA host galaxies using this technique yielded no confirmed detections in a sample of 32 DLAs, and the stack of all measurements yielded a SFR limit of $\lesssim0.3M_\odot$/yr within 2 kpc from the QSO position \citep[][]{Fumagalli:2015a}.

The {\ciis} technique is described in detail by \citet{Wolfe:2003a,Wolfe:2003b}. In short, the technique relies on equating the cooling rate dominated by [CII] 158 micron cooling with the heating rate, set in part by the star formation rate. By measuring the {\ciisl} transition, the cooling rate {\lc} can be calculated, which leads to an estimate of the SFR.
However, the inferred SFRs from this measured for typical DLAs by \citet{Wolfe:2008} may be in conflict with the results by \citet{Fumagalli:2015a}, as they predict higher SFRs in the typical DLA population. 
Careful calibration of the technique is required with a sample of galaxies with known SFRs. 
Since the {\ciisl}  absorption line arises from the same excited fine-structure state that gives rise to [C II] 158 {\micron}
emission, ALMA measurements of a sample of DLAs should help advance our understanding of {\ciisl}. 

The last technique, which is the one presented in this paper, relies on a statistical comparison to connect absorption line studies tracing the gas, with star-forming galaxies measured in emission. 
It does so by predicting the comoving SFR density from DLAs by combining the column-density distribution function of H\textsc{i} gas at high redshift \citep[e.g.][]{Prochaska:2009} with the 
Kennicutt-Schmidt (KS) relation \citep{Schmidt:1959, Kennicutt:1998a} and a geometrical model \citep{Rafelski:2011}, and then compares this prediction to emission that likely originates from DLAs.

Since the expected emission from DLAs is below 29 mag/arcsec$^2$ \citep{Wolfe:2006}, it is important to conduct any search for such emission in a very deep imaging field. 
We make this measurement in the Hubble Ultra Deep Field \citep[UDF;][]{Beckwith:2006}, because it is the only rest-frame UV imaging that has sufficient sensitivity and resolution to measure the SFR in emission at high redshift. 

There are two possible locations to observe emission from DLAs at high redshift; either in isolated regions unassociated with SFGs, or in the outskirts of SFGs. 
A search for such emission in isolated regions of the UDF was conducted at $z\sim3$ by \citet{Wolfe:2006}. They assumed that all the emission was unassociated with SFGs,
and found upper limits on the SFR efficiency of $\sim10$\%. This SFR efficiency is defined as the fractional decrease in the normalization of the KS relation, where $\sim10$\% is a factor of ten decrease in the normalization.

Since DLAs are expected to be associated with SFGs, a search for emission was conducted by \citet{Rafelski:2011} in the outskirts of $z\sim3$ SFGs in the UDF. 
In order to obtain reliable photometric redshifts of SFGs, the Lyman break of the spectral energy distribution (SED) needed to be sampled.  Galaxy redshifts were determined by both color selection and photometric redshifts including
ground based u-band data sampling the Lyman-break at $z\sim3$ \citep{Rafelski:2009}. The study found that if the emission in the outskirts of the SFGs is due to in-situ star formation (SF) in atomic-dominated hydrogen gas, then the SFR efficiency of the gas at $z\sim3$ is between $\sim2$ to 10\%. 

There are multiple possible effects contributing to the lower SFR efficiencies than predicted by the KS relation, such as the lower metallicity of DLA H\textsc{i} gas \citep[e.g.][]{Gnedin:2011}, 
a higher background radiation field at high redshift \citep[e.g.][]{Haardt:2012}, and the role of molecular versus atomic hydrogen gas in star 
formation \citep[e.g.][]{Glover:2012a,Krumholz:2012c, Krumholz:2013a}. Measurements of the evolution of the SFR efficiency and metallicity over time will help distinguish between these different possibilities. 

In order to measure the evolution in the SFR efficiency, we require accurate redshift estimates of SFGs. The newly observed UV imaging of the UDF \citep{Teplitz:2013} provide significantly improved redshifts at $z<4$ and the new redshift catalog reduces the outlier fraction by a factor of $\sim3$ \citep{Rafelski:2015}. These high fidelity redshifts and high resolution UDF images can then be used to create stacks of galaxies at varying epochs enabling the measurement of the SFR efficiency of atomic-dominated HI gas in the outskirts of SFGs at $z\sim1$ and $z\sim2$ while also improving the $z\sim3$ measurement

A comparison of the SFR of high redshift galaxies with the inferred molecular gas surface densities measured at $z\sim1-3$ suggest little to no evolution compared to the the KS relation measured at $z=0$ \citep{Daddi:2010,Tacconi:2010, Genzel:2010, Tacconi:2013}.  While these studies have measured the efficiency of star formation at high redshift in molecular-dominated gas, they do not address the SFR efficiency for neutral atomic-dominated H\textsc{i} gas in the galaxy outskirts, as done in this study.

The outline of this paper is as follows. In Section \ref{data} we analyze the data; we describe how we measure SFRs (\ref{bands}), define our galaxy samples (\ref{sample}), create composite stacks (\ref{stacks}), extract radial surface brightness profiles (\ref{profiles}), and consider the effects of Ly$\alpha$ on the profiles (\ref{lya}). In Section \ref{sfrefficiency} we describe how we obtain the SFR efficiency; we determine the column density distribution function of DLAs (\ref{fofn}), measure the SFR efficiency (\ref{efficiency}), and visualize the efficiency on the KS plot (\ref{vis}). In Section \ref{cf} we compare the covering fraction of the outskirts of SFGs with atomic-dominated DLA gas (\ref{atomic}) and with molecular-dominated gas (\ref{molecular}). We then discuss the results in Section \ref{discussion}; we compare the results to those of \citet{Rafelski:2011} (\ref{comp}), examine the two different possible normalizations of the $z\sim1$ column density distribution function (\ref{k3}), compare the results to predictions from \citet{Krumholz:2013a} in Section \ref{models}, compare the results to the SFR efficiency of local H\textsc{i} gas in Section \ref{local}, compare to measurements from the double DLA technique in Section \ref{double}, and consider the effects of dust (\ref{dust}). We summarize and conclude in Section \ref{summary}. 

For consistency with past results, we continue to use the Salpeter initial mass function \citep[IMF;][]{Salpeter:1955} and the Kennicutt far-ultraviolet (FUV) SFR calibration \citep{Kennicutt:1998b} throughout this paper. To convert to the Kroupa IMF \citep{Kroupa:2003}, divide all SFRs by a factor of 1.59. 
We adopt the AB magnitude system and an $(\Omega_M, \Omega_\Lambda, h)=(0.3,0.7,0.7)$ cosmology.

\section{Data Analysis}
\label{data}

The Hubble UDF ($\alpha(J2000) = 03^{h}32^{m} 39^{s}$, $\delta(J2000) = -27^{\circ}$47$\tt'$29.$\tt''$$1$) includes the most sensitive and high-resolution images of any part of the sky. We restrict ourselves to this field due to the extreme faintness of the expected emission from DLAs. Even with this exquisite sensitivity, the signal-to-noise in the outskirts of individual high-redshift SFGs is insufficient to directly measure the spatially extended star formation from atomic-dominated H\textsc{i} gas. We therefore create composite stacks for this measurement. In this section, we describe the bandpass selection for measuring SFRs in Section \ref{bands}, sample selection in Section \ref{sample}, creation of the composite stacks in Section \ref{stacks},  and extraction of the radial surface bright profiles to measure emission in the outskirts of SFG galaxies in Section \ref{profiles}.  We also consider possible contamination from Ly$\alpha$ in Section \ref{lya}. We use a procedure similar to that used in \citet{Rafelski:2011}, where we successfully measured emission in the outskirts of galaxies at $z\sim3$.

\subsection{Star Formation Rates}
\label{bands}

We determine spatially extended star formation around high redshift SFGs by measuring their rest-frame UV flux. The UV is a sensitive measure of the SFR since the UV photons are produced by short-lived massive stars. We convert the measured flux to a SFR by the calibration from \citet{Kennicutt:1998b}. We use three of the original ACS optical bandpasses, $B$, $V$, and $i^\prime$ (F435W, F606W, and F775W),  corresponding to the rest-frame FUV and NUV fluxes. Only the rest-frame FUV data near 1500\AA ~were considered by \citet{Rafelski:2011}, and this light falls in the $B$ and $V$ bands at $z\sim2$ and $z\sim3$. However, the UV spectrum is nearly flat from 1500-2800\AA~ \citep{Kennicutt:1998b}, which allows measurements of the SFR from two independent bandpasses at $z\sim2$ and $z\sim3$ ($V$ and $i^\prime$ bands), and it enables measurement of the $z\sim1$ sample in the optical B-band.  While we do have images at 1500\AA~ for the $z\sim1$ bin, the wavelengths fall in the WFC3/UVIS bandpasses. The UV images are less sensitive than the optical UDF images, and potentially suffer from charge transfer inefficiencies \citep{Teplitz:2013, Rafelski:2015}. 

\begin{figure}[]
\center{
\includegraphics[scale=0.5, viewport=10 10 500 340,clip]{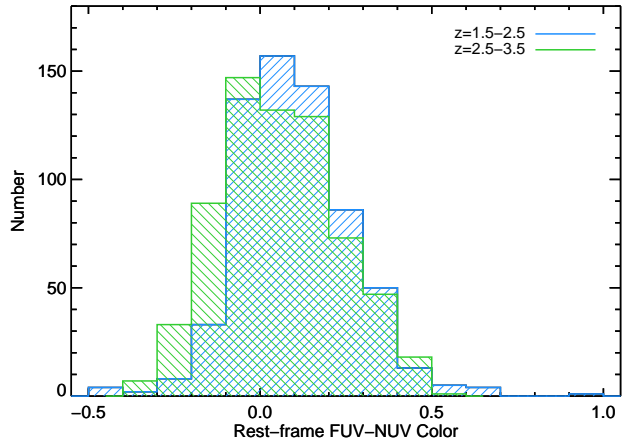}
}
\caption{ \label{fig:uvcolor} 
FUV-NUV rest-frame color for the full $z\sim2$ and $z\sim3$ samples. The median FUV-NUV color is close to 0, justifying the use of both the FUV and NUV images in measuring the SFR.}
\end{figure}

The FUV and NUV SFR calibrations by \citet{Kennicutt:2012} are very similar, which further supports using the FUV and NUV images as independent SFR measurements, assuming a continuous star formation history (SFH). We test whether it is equivalent to measure the SFR from anywhere within the 1500-2800\AA ~window by examining the FUV-NUV color for the full sample\footnote{See Section \ref{sample} for the definition of what the `full' sample is.} of SFGs in Figure \ref{fig:uvcolor}. We find a median FUV-NUV color of $0.09\pm0.16$ mag at $z\sim2$ and $0.04\pm0.17$ mag at $z\sim3$, which suggests that either the FUV or the NUV flux will yield the same SFR for the majority of our sample. The width of the distribution is larger than our uncertainties, with a slightly positive spread. This may be due to dust extinction of our galaxies, since the FUV flux is lower than the NUV flux.

The rest-frame FUV light from normal high-redshift SFGs suffers from dust extinction by up to a factor of 5 \citep{Reddy:2012a}, which could reduce our measured SFR efficiency later in this paper. However, the dust is concentrated in the center of the galaxies in the highest star-forming region and is much reduced for lower mass galaxies \citep{Nelson:2016}. Similarly, \citet{Bigiel:2010b} find that the FUV emission in the outskirts of local galaxies reflects the recently formed stars without large biases from external extinction. We similarly do not expect much extinction in the outer parts of SFGs, if it consists of atomic-dominated H\textsc{i} gas, as DLAs have low dust-to-gas ratios \citep{Murphy:2004, Frank:2010, Khare:2012, Fukugita:2015, Murphy:2016}. We therefore do not apply a dust correction to the SFRs, as the measured SFRs presumably occur in DLAs. However, we consider possible dust extinction further in Section \ref{dust}.

We also note that the SFR calibrations are sensitive to metallicity, and decreasing the metallicity should increase the FUV luminosity for a given mass distribution, and thereby the SFR \citep{Kennicutt:2012}. Since DLAs have a metallicity of approximately a thirtieth solar \citep{Rafelski:2012}, the SFRs of DLA gas are potentially underestimated by $\sim0.1$ dex.  However, the change from this effect is small given that the evolution in metallicity from $z\sim1$ to $z\sim3$ is $\sim0.4$ dex, and thus we do not consider it here.

\subsection{Samples}
\label{sample}

\begin{deluxetable*}{cccccccccccccccc}[h!]
\tablecaption{Sample properties \label{tab:samp}}
\tablehead{
\colhead{Sample} &
\colhead{$z_{\rm min}$} &
\colhead{$z_{\rm max}$} &
\colhead{Full samp} &
\colhead{Sub samp} &
\colhead{Scale} &
\colhead{Volume} &
\colhead{DA} \\
\colhead{} &
\colhead{} &
\colhead{} &
\colhead{} &
\colhead{} &
\colhead{kpc/arcsec} &
\colhead{Mpc$^3$} &
\colhead{Mpc}
}
\startdata
$z\sim1$ & 0.7 & 1.5 & 802 & 36 & 8.01 & 24555 & 1652 \\
$z\sim2$ & 1.5 & 2.5 & 1329 & 30 & 8.37 & 36914 & 1727 \\
$z\sim3$ & 2.5 & 3.5 & 1178 & 43 & 7.70 & 37297 & 1589
\enddata
\tablecomments{Redshift sample properties and adopted physical constants.Volume is the co-moving volume sampled. DA is the angular diameter distance. The full sample is significantly larger than the sub sample, which is due to a different magnitude cut, photometric redshift quality, and morphological selection. }
\end{deluxetable*}

We select galaxies using the photometric redshifts from \citet{Rafelski:2015}, which include the eleven Hubble Space Telescope bandpasses covering the UDF from the near-ultraviolet (NUV) to the near-infrared (NIR) \citep{Beckwith:2006, Oesch:2010a,Oesch:2010b,Bouwens:2011b, Koekemoer:2013, Ellis:2013, Illingworth:2013, Teplitz:2013}. The photometric redshift fits all cover either the Lyman-break or the 4000\AA ~break of the galaxies, and most cover both breaks. This yields robust redshift estimates, as evidenced by comparisons to spectroscopic and grism redshifts \citep{Rafelski:2015}. We emphasize that these redshift estimates are significantly better than those previously available, and this analysis would not be possible without the new NUV and NIR data.

In addition to the redshift selection, we require that the SED template from the photometric fit is a star-forming galaxy, and since most galaxies at $z>0.7$ are star-forming \citep{Muzzin:2013}, this is a small reduction in sample size (8\% at $z\sim1$, down to 2\% at $z\sim3$). We also impose a $V<29$ magnitude cut to help secure a sample with robust photometric redshifts, as the NUV images and the corresponding CANDELS \citep{Grogin:2011, Koekemoer:2011} portions of the NIR images have $5-\sigma$ depth of $\sim28-29$ mag \citep{Rafelski:2015}. We split our sample into three redshift bins, $z\sim1$, $z\sim2$, and $z\sim3$, as described in Table \ref{tab:samp}. These redshift bins enable us to sample the rest-frame FUV light in the most sensitive optical bandpasses, and provide sufficient galaxies per bin for a robust stack. This set of selection criteria will be considered the `full' sample per redshift bin, while other selection criteria below will create the subsamples for stacking. We note that we do not put a requirement on the quality of the photometric redshift in this full sample, since its purpose is to provide the total number of galaxies at that redshift down to $V<29$. This sample differs from the sample defined below, as it is meant to capture these galaxies down to a fainter magnitude than will be used for stacking. This is necessary for the completeness corrections discussed in Section \ref{measure}, and we clarify which sample is used as necessary.

\begin{figure*}[]
\center{
\includegraphics[scale=0.3, viewport=0 0 412 610,clip]{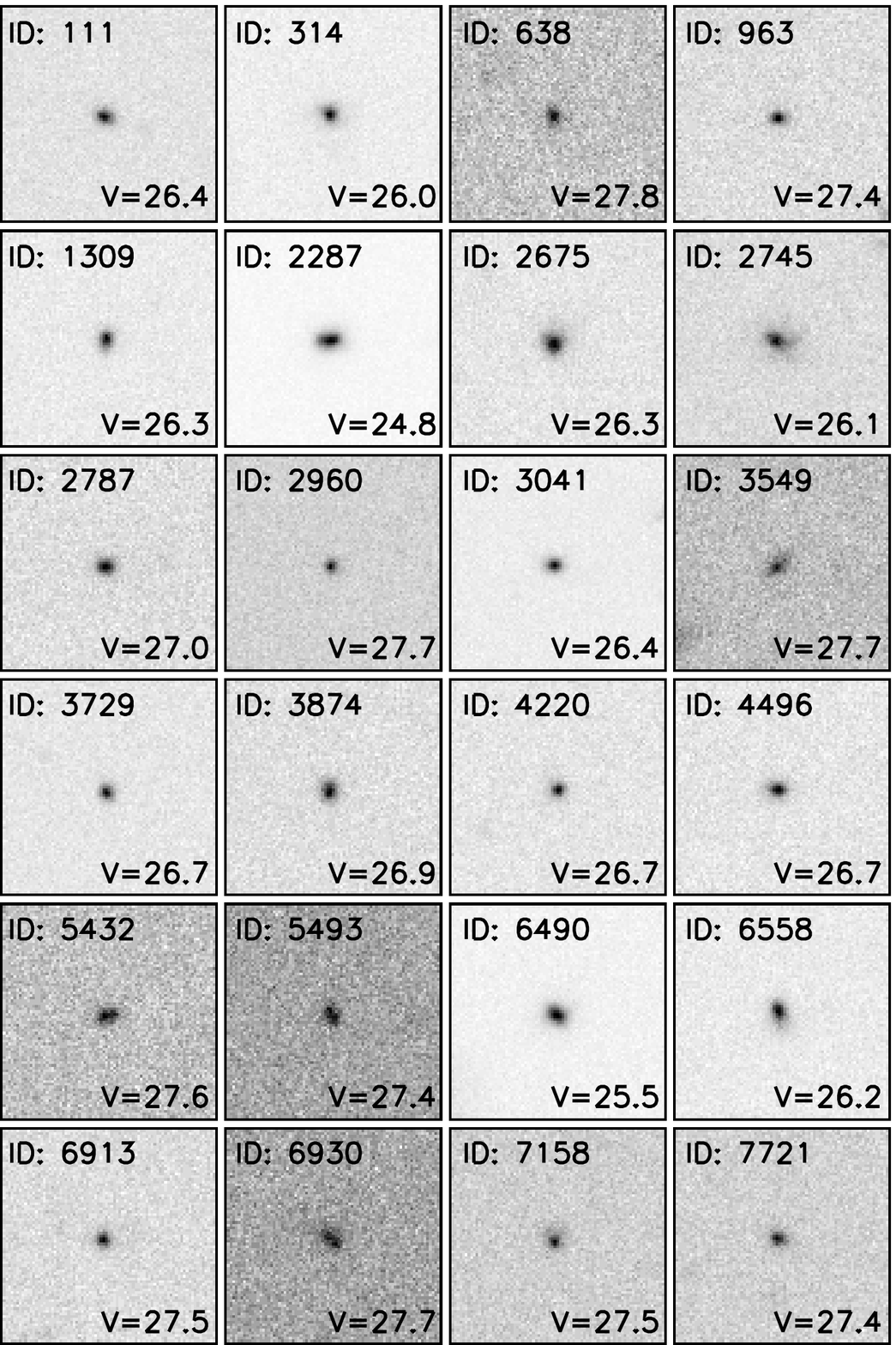}
\hspace{4mm}
\includegraphics[scale=0.3, viewport=0 0 412 610,clip]{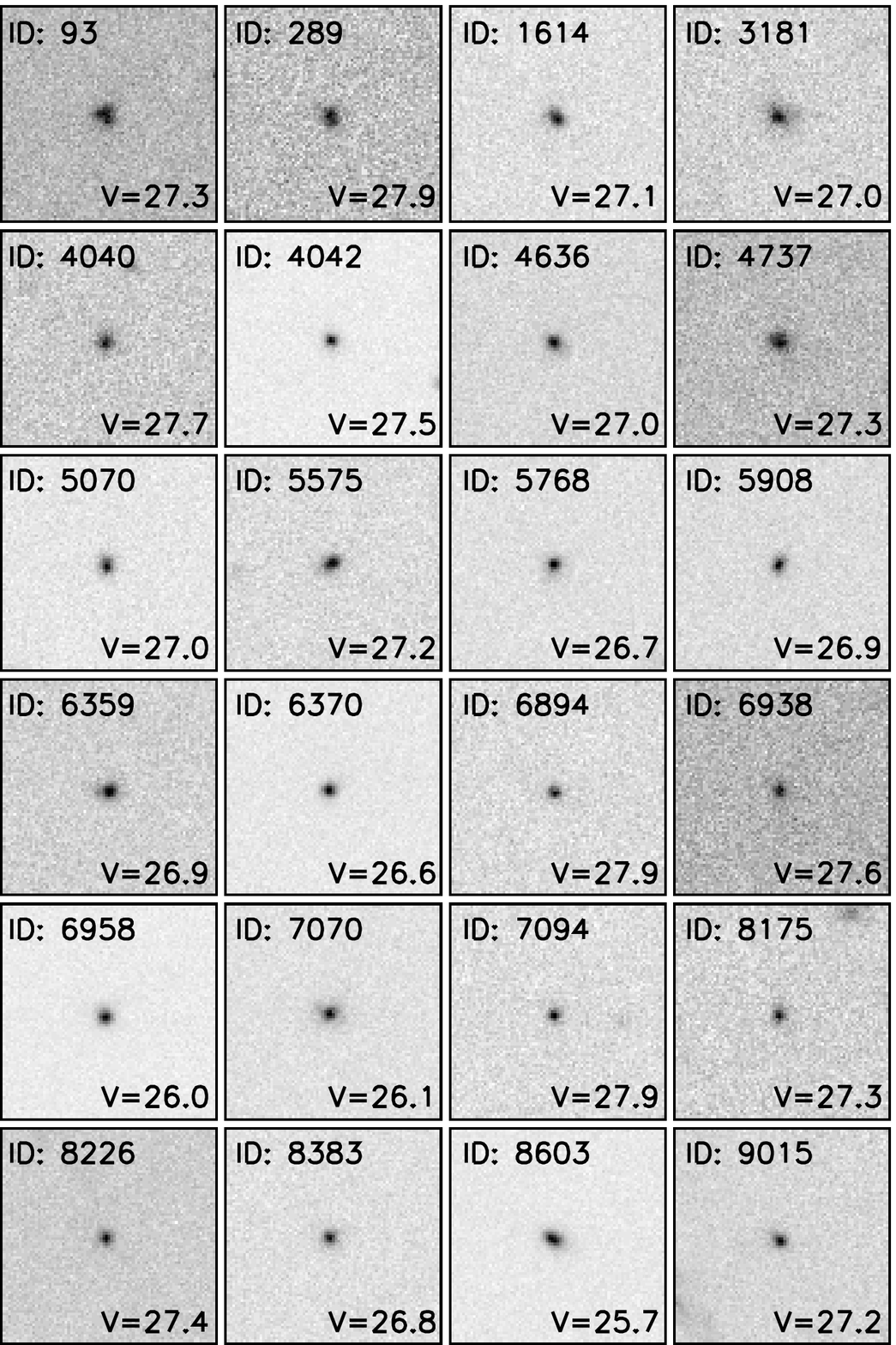}
\hspace{4mm}
\includegraphics[scale=0.3, viewport=0 0 412 610,clip]{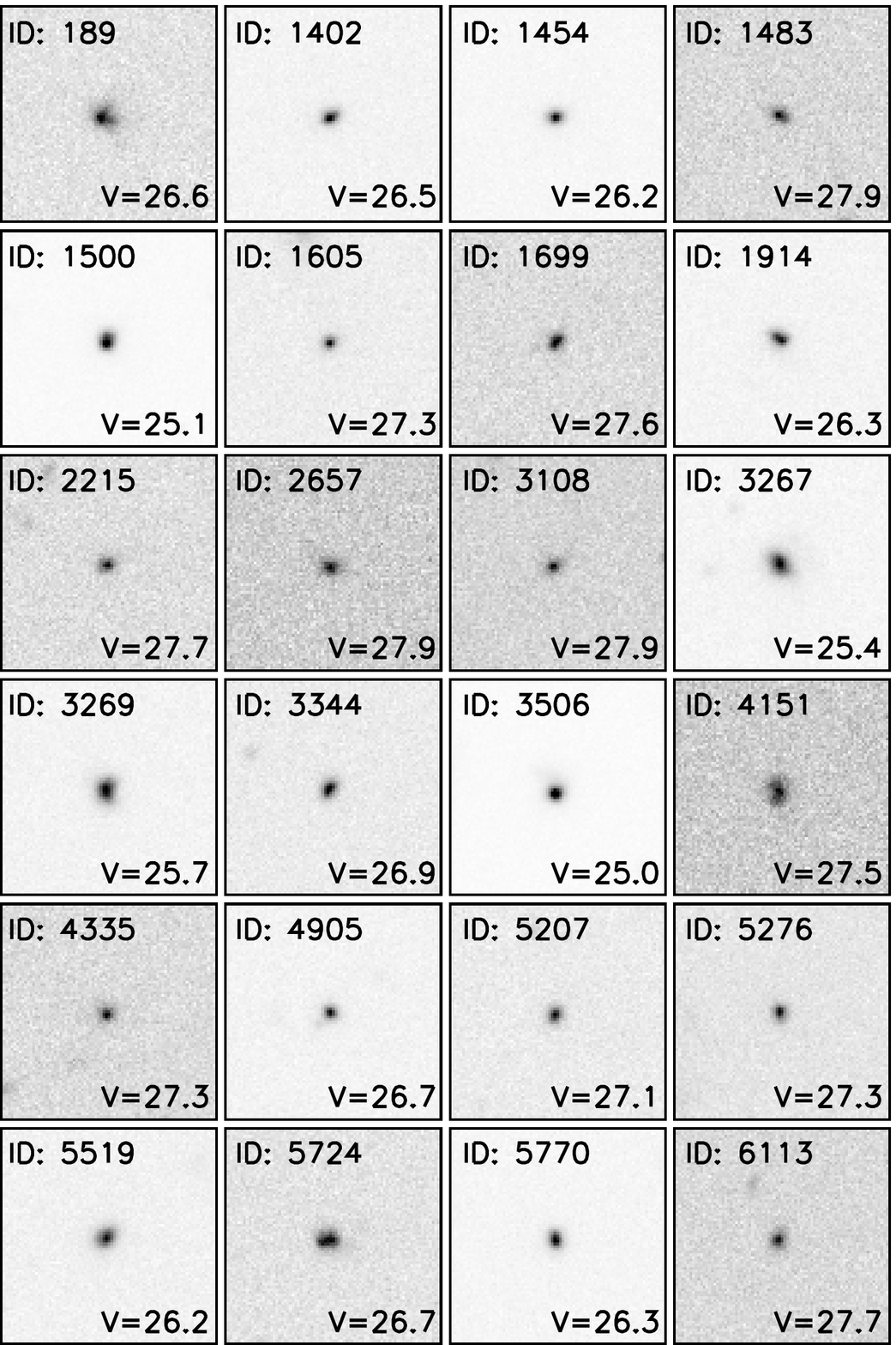}

\includegraphics[scale=0.3, viewport=0 100 412 610,clip]{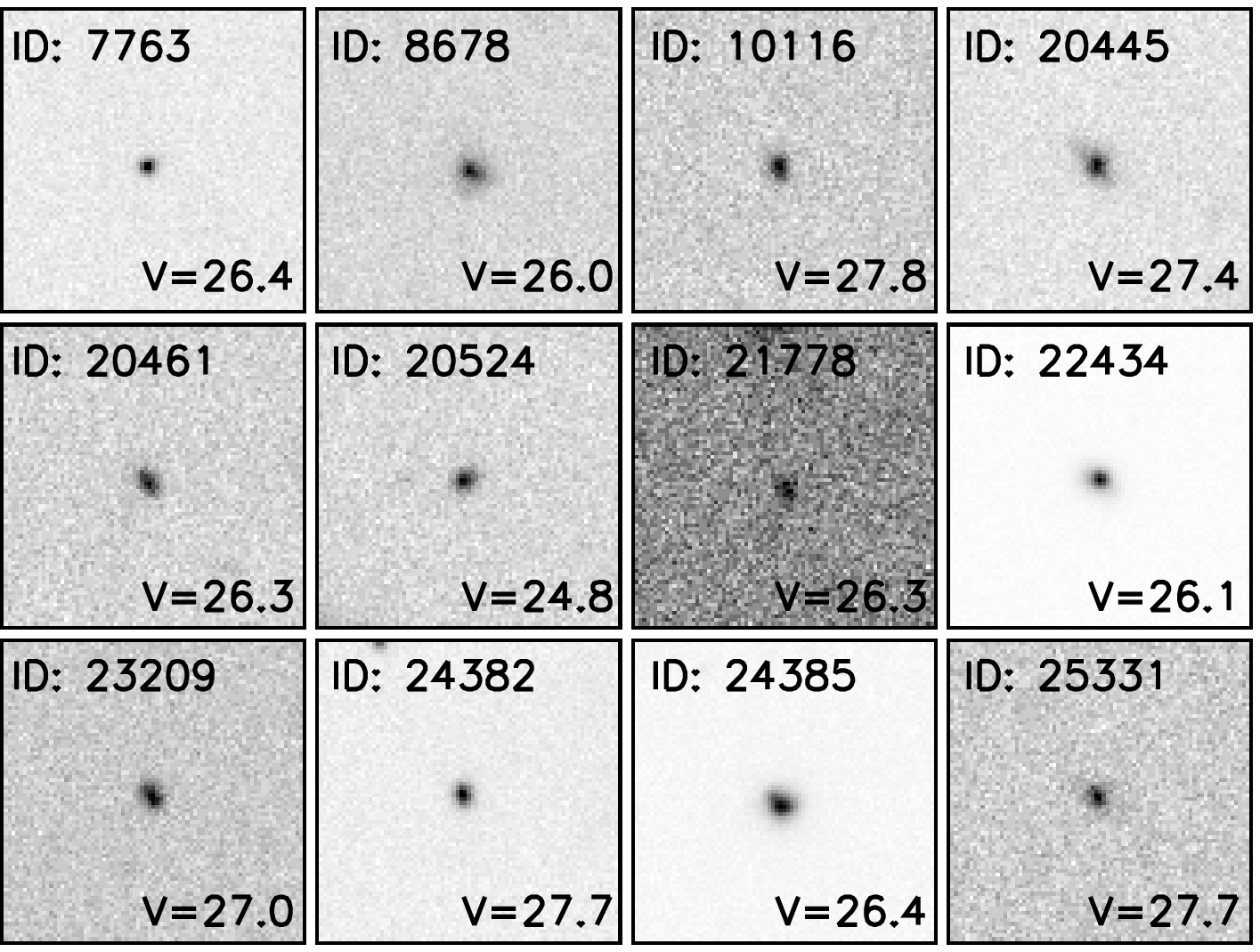}
\hspace{4mm}
\includegraphics[scale=0.3, viewport=0 100 412 610,clip]{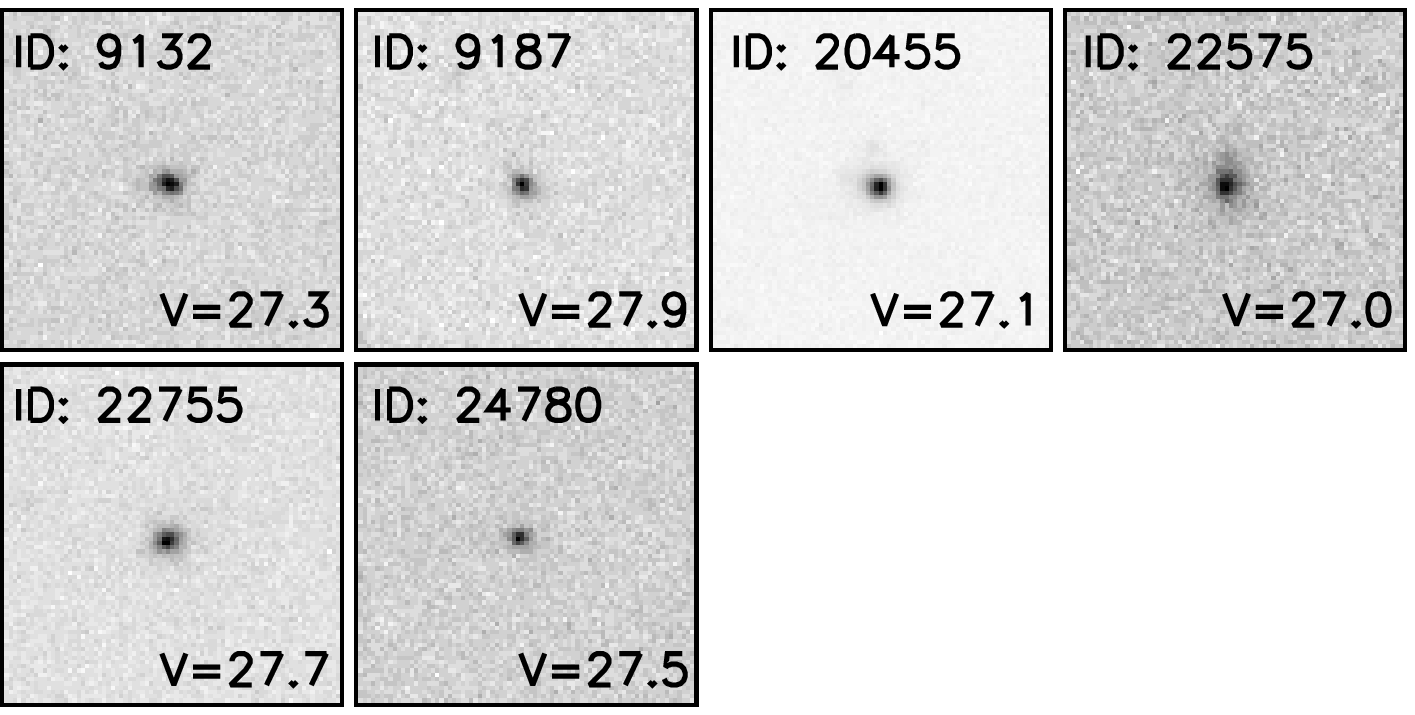}
\hspace{4mm}
\includegraphics[scale=0.3, viewport=0 100 412 610,clip]{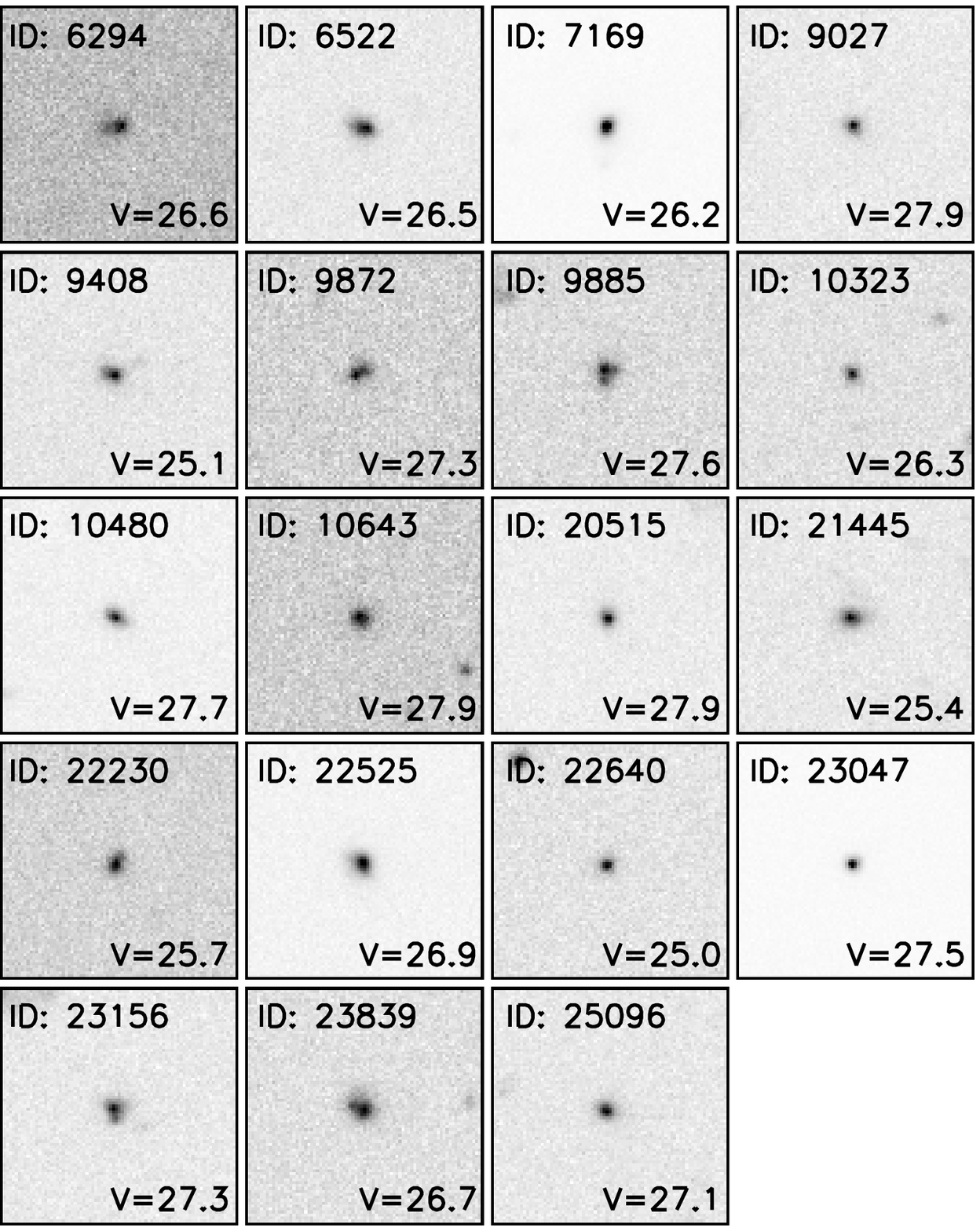}

}
\caption{ \label{fig:thumbnails} 
Thumbnail images at $z\sim1$ (left), $z\sim2$ (middle) and $z\sim3$ (right) of the subsample of galaxies used for stacking. The $z\sim1$ and $z\sim2$ thumbnails are from the F435W image and the $z\sim3$ sample is from the F606W image. This corresponds to the NUV at $z\sim1$ and FUV at $z\sim2$ and $z\sim3$. The thumbnails are 2.4 arcsec on a side, with F606W magnitudes and IDs from \citet{Rafelski:2015} labeled on each thumbnail. We note that this is a larger region than is extracted below, and shows that the galaxies are isolated. }
\end{figure*}

To create a composite stack of SFGs in each redshift bin, we select a sample with similar morphological and physical characteristics, and without nearby neighbors that can overlap the galaxies or cause dynamical disturbances. We therefore select compact, symmetric, and isolated galaxies in a similar fashion as \citet{Rafelski:2011} and \citet{Hathi:2008}. Morphological parameters are determined with $\tt SExtractor$ \citep{Bertin:1996} on the V-band image, and selected using the full width half maximum (FWHM) for compactness and ellipticity $\epsilon = (1-b/a)$ for circularity. We require a FWHM $\leq$ 2 kpc, $\epsilon \leq0.25$, and no nearby objects brighter than 29th magnitude within 11 kpc. In addition, we visually inspect the galaxies, and require that no nearby companion is visible, but would have just passed the magnitude cut. This enables us to maximize the sample size without any visible contaminants. We limit ourselves to galaxies with very good photometric redshift fits, with $\tt ODDS>0.9$ \citep{Benitez:2000} and $\chi^2<4$, as this was found to produce reliable photometric redshift selected samples \citep{Rafelski:2009, Rafelski:2015}.  Furthermore, we impose a $V<28$ magnitude cut to ensure that the individual galaxies have sufficient S/N in the optical images for morphological analysis, which also results in improved photometric redshifts for the stacked sample. These are basically the same criteria as by \citet{Rafelski:2011}, modified to be based on a physical distance for consistency between redshift bins. Table \ref{tab:samp} lists the  parameters for the various subsamples, including the scale used to convert to arcseconds. Thumbnails of the galaxies in each subsample to be stacked are shown in Figure \ref{fig:thumbnails}.

The $z\sim3$ redshift bin is the same as in \citet{Rafelski:2011}, but the improved photometric redshifts cause different galaxies to be selected. Of the 48 galaxies selected by \citet{Rafelski:2011}, 35 have new photometric redshifts in the range $2.5<z<3.5$. For the remaining 13 galaxies with different redshifts, 2 are catastrophically incorrect ($z<1$), 5 include the redshift bin within their uncertainties, and the other 6 have a median redshift of $z\sim2.3$.  Although the \citet{Rafelski:2011} study includes some galaxies with incorrect redshifts, the stacks are median values and thus are relatively robust against such systematics. Therefore, the results presented by \citet{Rafelski:2011} are still valid. In Section \ref{comp} we find excellent agreement between our new results and \citet{Rafelski:2011}. 

We compare the magnitude, color, and redshifts of the full and sub-samples to check for any differences in the properties of the two samples.  The magnitudes are measured in the F606W bandpass, the colors based on the F435W-F160W, F435W-F850LP, and F105W-F160W colors, and the redshifts using the new photometric redshifts. 
Since the full and sub-samples apply different magnitude cuts, we remove the additional magnitude cut for the sub-sample in this comparison, as we are interested in determining if the morphological and redshifts cuts result in any changes in the sample properties.

We find the median values of magnitudes, colors, and redshifts are similar, with no systematic trends. We apply the Kolmogorov-Smirnov test, and find that all three distributions are consistent with being drawn from the same parent population. The fact that the magnitudes are similar suggest that the SFR of the two samples are similar. The similar distribution of colors suggest that the two samples are made up of the same stellar populations and have similar star formation histories.

\subsection{Composite Stacks}
\label{stacks}
For each redshift sample, we create a galaxy stack by centering on each galaxy with a Gaussian fit to sub-pixel precision and shifting to a common reference grid with a damped sinc function, and then obtain the median of the images. This creates a stack which is insensitive to outliers, and therefore any single galaxy at an incorrect redshift, with an unidentified nearby neighbor, or with extreme Ly$\alpha$ emission would not significantly affect the stack. In \citet{Rafelski:2011} we tested this stacking procedure and found it robust to varying brightness, color, or FWHM within a stack. We also tested the difference in taking the median and the mean, and although the central regions of the mean stacks where higher, we found little difference in the galaxy outskirts.  \citet{Rafelski:2011} therefore chose to use the median value rather than the mean, given that it is more robust to contamination or incorrect photometric redshifts, and we similarly use the median value here. We note that any minor potential biases from asymmetric star formation would be included in all the redshift bins, and therefore not affect the evolution measurement.

\begin{figure}[]
\center{
\includegraphics[scale=0.47, viewport=0 0 530 175,clip]{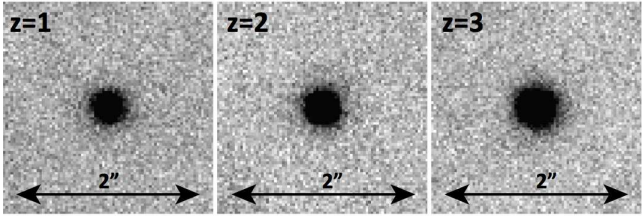}
}
\caption{ \label{fig:stacks} 
Images of galaxy stacks at $z\sim1$, $z\sim2$ and $z\sim3$, each 2 arcseconds wide and stacked in $\mu$Jy to avoid zeropoint differences. The $z\sim1$ stack includes the B-band, the $z\sim2$ stack includes B and V-bands, and $z\sim3$ stack includes V and i$^\prime$-band. }
\end{figure}

At $z\sim2$ and $z\sim3$ we include both the rest-frame FUV and NUV data within a single stack to improve the S/N, and thus stack in $\mu$Jy rather than electrons due to the different zeropoints of the images. The $z\sim2$ stack consists of the F435W and F606W images and the $z\sim3$ stack consists of the F606W and F775W images. For the $z\sim2$ and $z\sim3$ samples we also create stacks just in a single passband in the FUV and NUV, and find them to have consistent radial surface brightness profiles. We choose to use the combined stacks to have the best signal-to-noise possible. The galaxy stacks are shown in Figure \ref{fig:stacks}.

As a comparison sample, we also create stacks of stars in the same fashion based on the sample by \citet{Pirzkal:2005}. We only include stars that are not saturated in each passband, isolated from nearby galaxies, and with confirmed grism spectra. This leaves us with a sample of 12 stars. Since we wish to determine the shape of the star's PSF rather than the absolute normalization, and because the stars have a wider distribution of magnitudes, we scale each star by its peak in a two-dimensional Gaussian fit to improve the PSF determination. We do not scale the galaxy stacks since in those stacks the absolute measured flux value is important, and the galaxies have a smaller magnitude range. Also, for the $z\sim2$ and $z\sim3$ galaxy stacks, we create a combined PSF including stars in both bands. 

\subsection{Radial Surface Brightness Profiles}
\label{profiles}

\begin{figure}[b!]
\center{
\includegraphics[scale=0.5, viewport=10 10 470 340,clip]{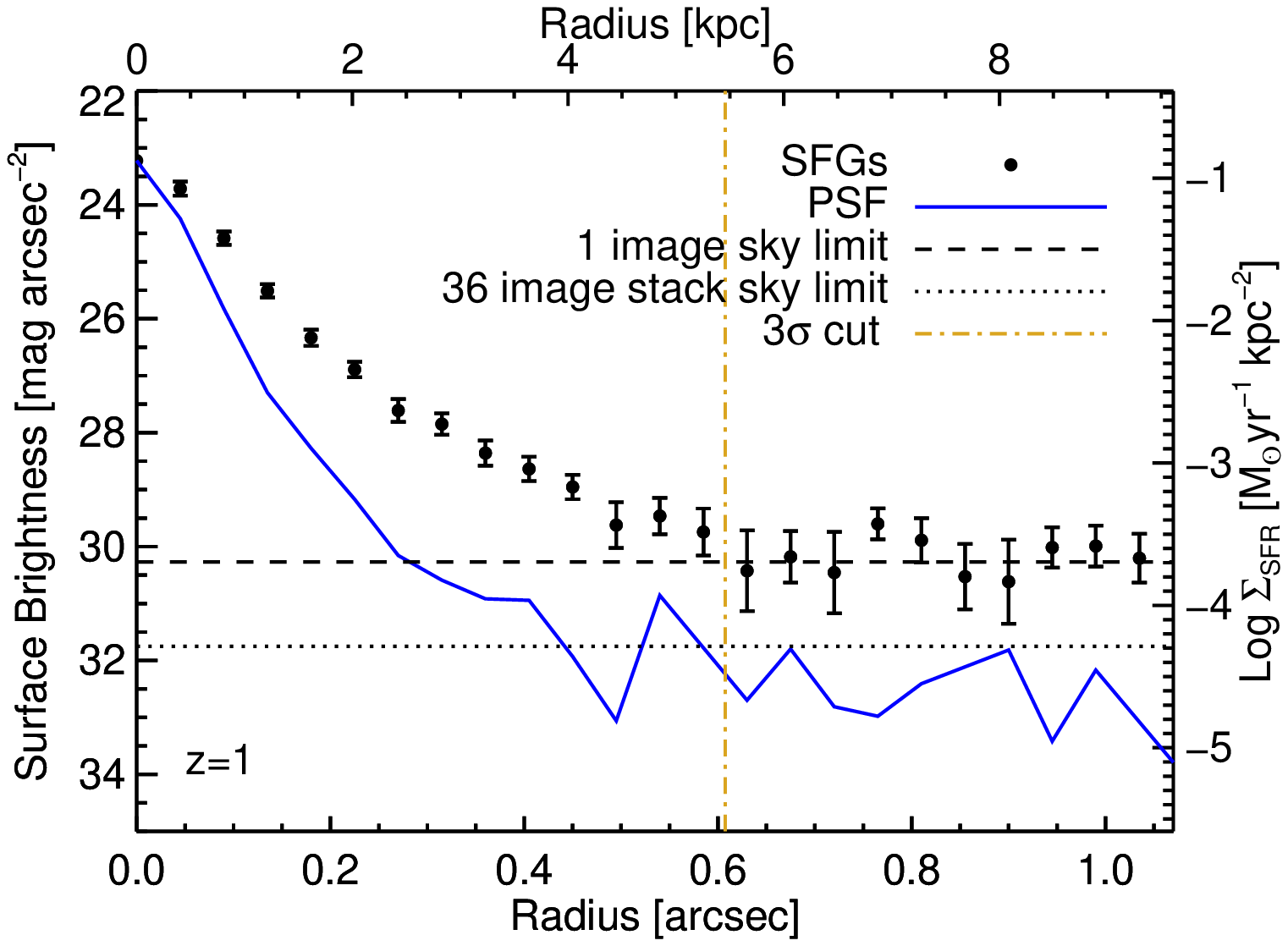}
\includegraphics[scale=0.5, viewport=10 10 470 340,clip]{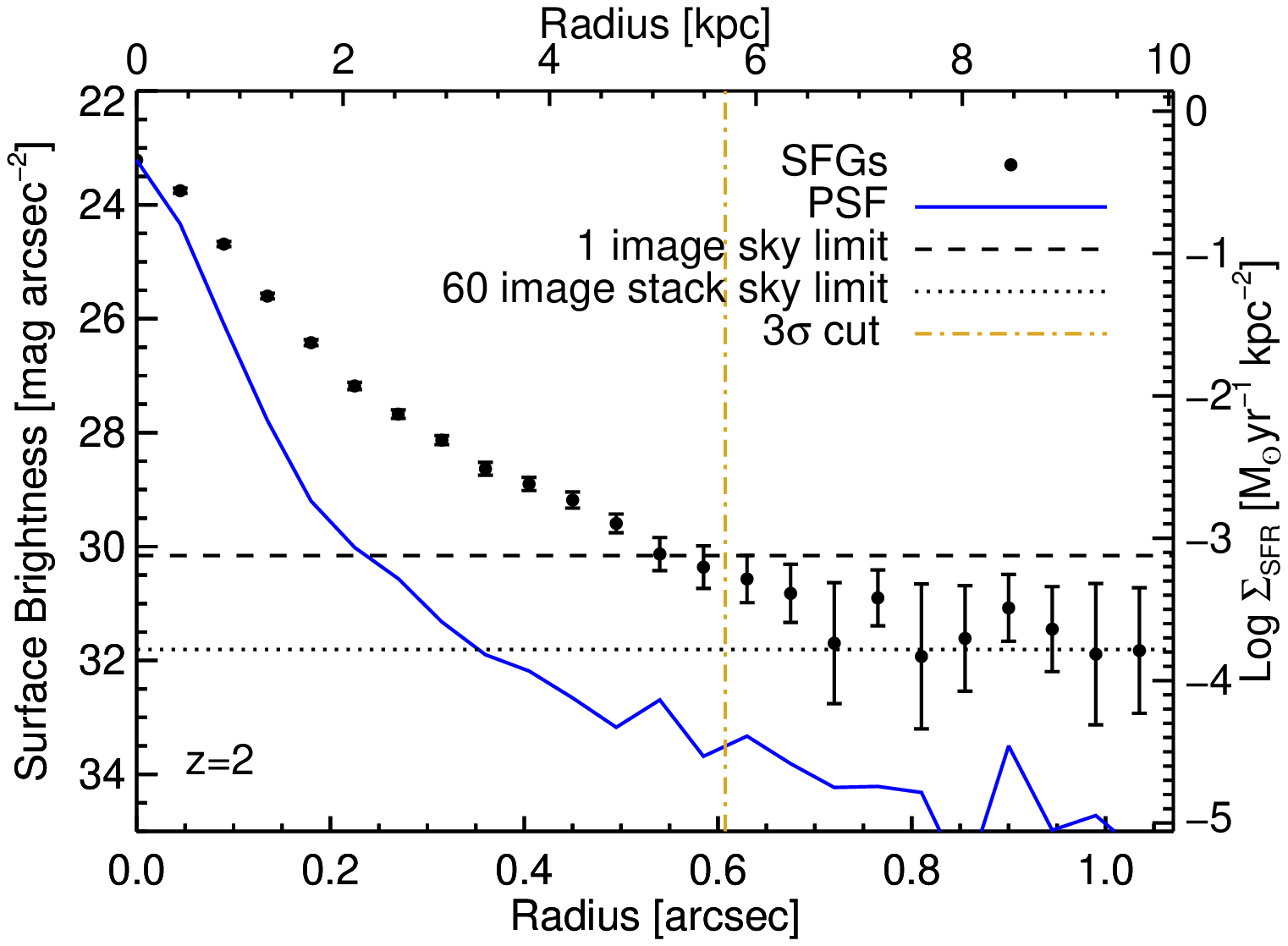}
\includegraphics[scale=0.5, viewport=10 10 470 340,clip]{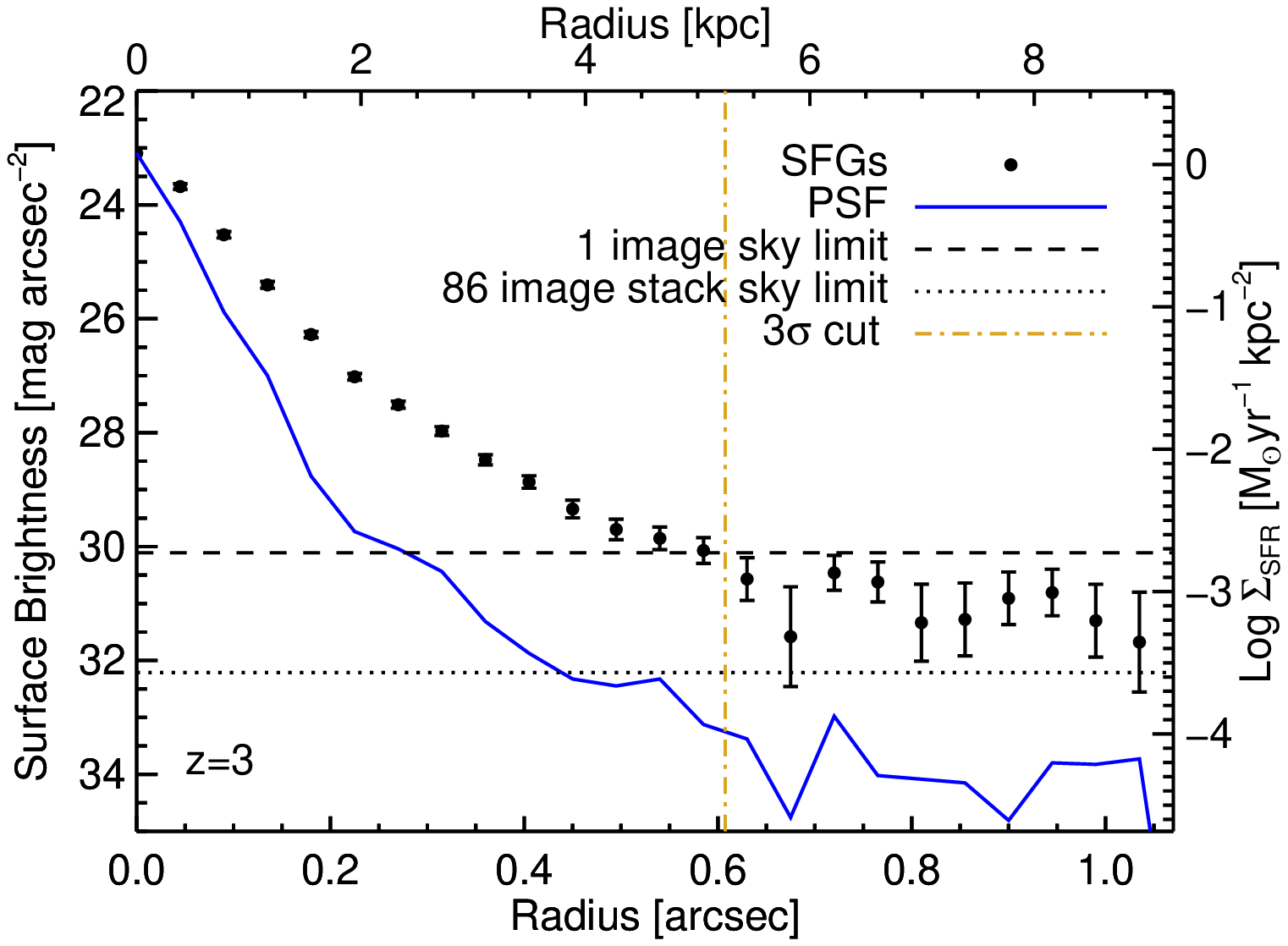}
}
\caption{ \label{fig:sbprofiles} 
Radial Surface Brightness profiles for the three composite stacks at $z\sim1$, $z\sim2$ and $z\sim3$ from Figure \ref{fig:stacks}. These profiles show extended star formation in the outskirts of SFG galaxies at high redshift, and are used in determining the SFR efficiencies below.The black points are the SFG profiles and the blue line is the PSF measured from stars. 
The dashed line is the 1-$\sigma$ sky uncertainty for a single galaxy, and the dotted line is the 1-$\sigma$ sky uncertainty for the stacked samples. Stacks of empty regions are at or below this dotted line. The uncertainties in the black points are a combination of the bootstrap uncertainty and the sky uncertainty. The gold dotted-dashed line corresponds to the 3-$\sigma$ cut used in Section \ref{measure}. 
}
\end{figure}

We extract radial surface brightness profiles from the three composite stacks in Figure \ref{fig:stacks} with the same methodology as by \citet{Rafelski:2011}. We use circular aperture rings with radial widths of 1.5 pixels, with no overlap, providing independent measurements at each radius. The star stacks are extracted in the same way, and then scaled to match the SFG stacks at the center. The extracted profile for each galaxy redshift bin and the relevant star profile is shown in Figure \ref{fig:sbprofiles}. The PSF declines more rapidly than the galaxy profile at all radii, which shows that the SFGs are resolved and have an extended profile. 

The uncertainty on each ring is determined by the same bootstrap analysis described by \citet{Rafelski:2011}, with a sampling of 1000 iterations. In addition to the sky uncertainty added in quadrature, this includes the uncertainty in sample variance, which accounts for possible contamination in the sample such as any catastrophic photometric redshift errors. The sky uncertainty is investigated by \citet{Rafelski:2011}, and we use the same procedure to obtain sky uncertainties here. 

Specifically, we determine the local sky background for all the SFGs by measuring the sky background in each thumbnail using an iterative rejection routine to discard flux from outlying pixels. The background pixels are then fit by a Gaussian to obtain 1-$\sigma$ sky values, which also confirms the sky level is at zero. The uncertainty in the sky decreases in the composite stack, and for a median stack in the Poisson limit it decreases by  $1.25 / \sqrt{N}$, where $N$ is the number of images. This was confirmed for the UDF by \citet{Rafelski:2011}, and we therefore apply this to the 1-$\sigma$ sky values to obtain the stacked limits. 

We further test the stacks by stacking empty regions of the sky and extracting their flux in the same fashion, and find the resulting flux levels are at or below this 1-$\sigma$ level shown as the dotted line in Figure \ref{fig:sbprofiles}, confirming no leftover residual flux from the sky. The 1-$\sigma$ levels are different in the different panels in Figure \ref{fig:sbprofiles} as they are determined in images at different wavelengths with different sensitivities.

The radial surface brightness depends on the sample selection described in Section \ref{sample}. Since the magnitudes, colors, and redshifts of the subsample and full sample are similar, we conclude there that the two samples have the same stellar populations and have similar star formation histories. Even so, the surface brightness profile is different if we stack the full sample or the subsample. This is investigated in Appendix B in \citet{Rafelski:2011}, and the result is that the full sample stack results in a surface brightness profile with a slightly elevated tail at larger radii. However, as noted in \citet{Rafelski:2011}, it is difficult to ascertain the amount of this emission that is caused by contamination from nearby neighbors or from morphologically different galaxies. If we ignore the contamination (which is significant), the SFR efficiencies determined in Section \ref{compare} would be increased by a factor of about two. This sets an upper limit on the any potential biases in sample selection on the radial surface brightness and the resultant SFR efficiencies. 

\subsection{Effects of Ly$\alpha$ on the surface brightness profiles}
\label{lya}

The Ly$\alpha$ emission line falls into the rest-frame FUV band-pass, and could potentially contaminate the measured flux. 
We consider if Ly$\alpha$ could be contaminating our stacks, and thereby the resulting surface brightness profiles. 
Previously, \citet{Rafelski:2011} conducted a test to see if a typical SFG Ly$\alpha$ line would significantly affect the measured photometry
by using the stacked Lyman break galaxy (LBG) spectrum from \citet{Shapley:2003} to estimate the effect, and found it to be negligible. However, it is possible that a subset of our sample have much stronger Ly$\alpha$ emission than a typical LBG, such as the green peas \citep{Henry:2015}, which could bias the SFRs obtained from the rest-frame FUV continuum high.

To test for this possibility, in particular in the galaxy outskirts, we conduct the following tests. First, for the $z\sim2$ and $z\sim3$ samples we create a smaller sample including about half of galaxies for which Ly$\alpha$ does not fall in the FUV bandpass by constraining the redshift, and compare the radial surface brightness profile of that stack to the full sub-sample stacks. We find no evidence of a decrease in the flux in the outskirts of the galaxies within their uncertainties. Second, we compare the radial surface brightness profiles for the FUV and NUV bands, since the NUV bands will not include the Ly$\alpha$ line. Again, we find no evidence of a change in the radial surface brightness profile in the outskirts. 

The fact that the galaxies have the same radial surface brightness profile shapes in both bands suggests that the Ly$\alpha$ line has little to no effect on the radial surface brightness profiles. We note that this is not in conflict with the results by \citet{Steidel:2011}, who find extended Ly$\alpha$ halos around SFGs.  First, that study samples a different galaxy population of brighter SFGs. Second, the \citet{Steidel:2011} study goes out to $\sim$80 kpc using ground based observations, while we limit ourselves to the central $\sim$10 kpc which is barely resolved in that ground-based study. Third, \citet{Steidel:2011} use narrowband filters to find the Ly$\alpha$, while our stacks are in very broad filters, and thus the Ly$\alpha$ would not contribute except in the extreme emitter cases, and then only in the FUV bands for part of the redshift range.

\vspace{5mm}

\section{Star formation rate efficiency}
\label{sfrefficiency}

The radial surface brightness profiles of the SFGs show spatially extended star formation in the galaxy outskirts. Star formation is expected in the outskirts of high redshift galaxies, as it is measured in atomic-dominated hydrogen gas at low-redshift \citep{Thilker:2007, Boissier:2008, Fumagalli:2008, Bigiel:2010a, Bigiel:2010b, Elmegreen:2015}. Such emission was also previously found at $z\sim3$ by \citet{Rafelski:2011} and interpreted as in-situ star formation in atomic-dominated hydrogen at high redshift. 
In this paper we will work under the hypothesis that the observed emission in the SFG outskirts is from in-situ star formation in atomic-dominated gas. We later consider this hypothesis in Section \ref{cf} by investigating the covering fraction of both atomic and molecular hydrogen gas compared to the observed emission. 

In this section we aim to measure the evolution in the SFR efficiency from $z\sim1$ to $z\sim3$ using the radial surface brightness profiles of the SFGs. In Section \ref{fofn} we determine the evolution in the normalization of the column-density distribution function of atomic-dominated neutral H\textsc{i} gas, $f(N_{\rm HI})$, and in Section \ref{efficiency} we determine the efficiency. In Section \ref{vis} we transfer the results to a KS relation plot to visualize the efficiency and compare with other studies and models in Section \ref{discussion}.

\subsection{Column-density distribution function}
\label{fofn}

The SFR efficiency and covering fraction of the atomic-dominated neutral H\textsc{i} gas (DLAs) directly depends on $f(N_{\rm HI})$. The $f(N_{\rm HI})$ is generally modeled with a double power-law to the number of high density absorbers to background quasars in surveys such as the Sloan Digital Sky Survey \citep[e.g.][]{Ahn:2012}. It takes the form:

\begin{equation}
f(N_{{\rm H  I}},X)=k_{3}\left(\frac{N}{N_{d}}\right)^{\alpha}\;,
\hspace{2mm}
\alpha = \left\{\begin{array}{ll}
\alpha_{3}  \ ; N \leq N_d  \;,\\
\alpha_{4} \ ; N > N_d  \;,\\
\end{array}
\right.
\label{eq:fofN}
\end{equation}

\noindent where  $k_{3}$ represents the normalization, $\alpha$ the slope, and $N_d$\footnote{Throughout this paper, all column densities are in log 10 with units of cm$^{-2}$.}
the break column density ($N_{\rm HI}$) in the double power-law. The exact values of these vary depending on the analysis 
technique, redshift range, and survey \citep{Prochaska:2005, Prochaska:2009, Noterdaeme:2009, Noterdaeme:2012}. 
The slope of $f(N_{\rm HI})$ is not observed to evolve with time at any column-density, but the normalization of $f(N_{\rm HI})$ does show evolution \citep[][]{Prochaska:2009,SanchezRamirez:2015}. 

For consistency with \citet{Rafelski:2011}, we adopt  $\alpha_{3}$=$-$2.0 for $N\le N_d$ \citep{Prochaska:2009}, 
and $\alpha_{4}$=$-$3.0 for $N>N_d$. The value of $\alpha_{4}$ is less certain than $\alpha_{3}$, and 
the value used is consistent both with the formulation of randomly-oriented disks used below, and 
with the value measured by \citet{Noterdaeme:2009}.

With the value of the slopes set, we determine the other parameter values by fitting the data presented by \citet{Noterdaeme:2012}, including
their completeness and systematic corrections\footnote{\citet{Noterdaeme:2012} do not explicitly measure $f(N_{\rm HI})$ as a function of redshift, just the cosmological mass density, which is the integrated quantity.}. First, we fit all the data simultaneously over the entire redshift range with ${\rm log} ~N_{\rm HI}>$20.3,
holding the slopes constant at the above values, and find log $N_d$=21.51. We note that if we allow all parameters to be free, 
then we recover similar slope values as defined above. 

\begin{figure}[]
\center{
\includegraphics[scale=0.47, viewport=0 0 500 350,clip]{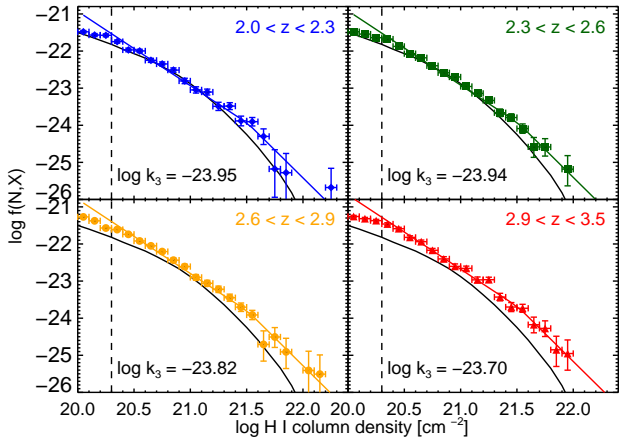}
}
\caption{ \label{fig:fofn} 
Fits to the column-density distribution function $f(N_{\rm HI})$ based on the data and corrections by \citet{Noterdaeme:2012} in 4 redshift bins; 
$2.0<z<2.3$, $2.3<z<2.6$, $2.6<z<2.9$, and $2.9<z<3.5$ in blue, green, orange, and red respectively. The data are shown as diamonds, squares, circles and triangles
respectively, and the fits as the solid lines in their respective colors. The fits are only for $N\ge$20.3, shown as a vertical dashed black line. 
We overplot the $z=0$ value of $f(N_{\rm HI})$ as determined from 21cm emission studies \citep{Zwaan:2005} for comparison, useful to visualize
the evolution in the normalization of $f(N_{\rm HI})$ in each redshift bin. 
A clear increase in the normalization of $f(N_{\rm HI})$ with redshift is observed relative to the z=0 value. 
}
\end{figure}

\begin{figure}[]
\center{
\includegraphics[scale=0.5, viewport=10 15 460 330,clip]{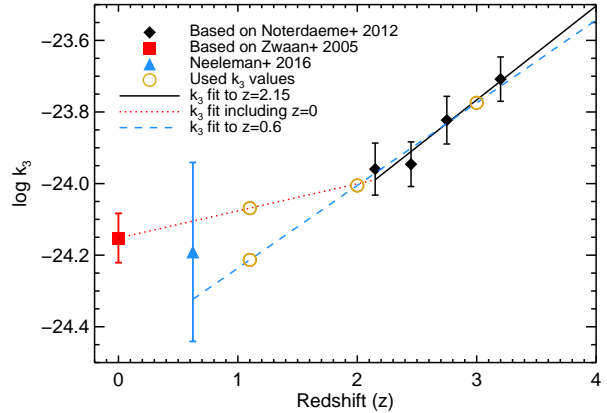}
}
\caption{ \label{fig:fofn_evol} 
Evolution in the normalization of $f(N_{\rm HI})$, $k_3$, as a function of redshift. The gold circles mark the adopted values
of $k_3$ used throughout this paper.
The solid black diamonds are from the fits in Figure \ref{fig:fofn}, and the black line is the best fit to these points.
The red dashed line is obtained by connecting the $z=0$ value with the $z=2$ value. 
The blue triangle is from \citet{Neeleman:2016a}, and the blue dashed line is a fit including the black diamonds and the blue triangle. }
\end{figure}

We then fix both the slopes and $N_d$, 
and split the data into 4 redshift bins. We use the same redshift bins as by \citet{Noterdaeme:2012}, except we combine the two highest
redshift bins to obtain better statistics.  The resultant fits in Figure \ref{fig:fofn} show an evolution in the normalization $k_3$ with redshift. 
We find that log $k_3=$ $-23.86$, $-23.94$, $-23.82$, and $-23.70$ for the redshift bins $2.0<z<2.3$, $2.3<z<2.6$, $2.6<z<2.9$, and $2.9<z<3.5$. 
We investigate the evolution of the normalization in Figure \ref{fig:fofn_evol}, which shows a clear increase in the normalization of $f(N_{\rm HI})$ with redshift.
This is consistent with the redshift evolution observed in the integral quantities (incident rate and mass density) of DLAs \citep{Noterdaeme:2012}.
We note that while the DR12 sample of DLAs is not yet publicly available, the internal DR11 sample appears to be in good agreement in the high $N_{\rm HI}$ regime \citep{Noterdaeme:2014}.

In addition to the redshift range directly sampled, we also require the value of the normalization at $z\sim1$. However, due to the atmospheric cutoff, 
it is difficult to observe DLAs at $z<2$, as it requires using space telescopes such as HST. Even if including all data obtained with HST, 
the resultant measure of the normalization of $f(N_{\rm HI})$ at $z=0.6$ has very large uncertainties \citet[][blue triangle]{Neeleman:2016a}. 
We therefore must obtain $k_3$ at $z\sim1$ using a fit dominated by the high redshift points, or with an interpolation including the $z$=0 value. 
The blue dashed line in Figure \ref{fig:fofn_evol} is a fit to the log $k_3$ values at $z>0.6$, and the black line is a fit to only the $z>2$ log $k_3$ values.
We adopt the resultant $k_3$ values from the blue dashed line, but note that an extrapolation of the $z>2$ fit to $z\sim1$ only differs from the full fit by 0.05 dex, due to the large uncertainties of the $z=0.6$ point.

The shape of $f(N_{\rm HI})$ at $z$=0 is different from the high redshift ones, shown as the solid black line in Figure \ref{fig:fofn} based on \citet{Zwaan:2005}.
This is likely due to the different method used to measure the H\textsc{i} gas, namely 21cm line emission rather than Ly$\alpha$ in absorption to background QSOs.
Moreover, the exact shape and normalization of the $z$=0 point is uncertain, with it depending on the dataset and method used to obtain $f(N_{\rm HI})$ \citep{Braun:2012}. For instance, \citet{Braun:2012} includes corrections at high column-densities, which significantly changes the slope. 
Therefore, while we can fit the same function to the $z=0$ data, it results in a poor fit due to the different (but uncertain) shape. However, the value of this fit is still useful to constrain possibilities for the value 
of $k_3$ at $z=1$, and we therefore show it as the red square in Figure \ref{fig:fofn_evol}, and connect it to the $z\sim2$ value with the red dotted line. This yields a 
second value for $z=1$ by interpolating in-between. We thereby have two possible values of $k_3$ at $z=1$, and consider both cases in what follows.
We note that the first value of $k_3$ is consistent with a near constant evolution between $z\sim2$ and $z\sim1$, whereas the second case is consistent with little to no evolution, and therefore the true $k_3$ value at $z\sim1$ 
is likely bracketed by these two measurements. Figure \ref{fig:fofn_evol} shows the four values of $k_3$ adopted in this paper as gold open circles.

\begin{figure*}[]
\center{
\includegraphics[scale=0.4, viewport=10 5 490 340,clip]{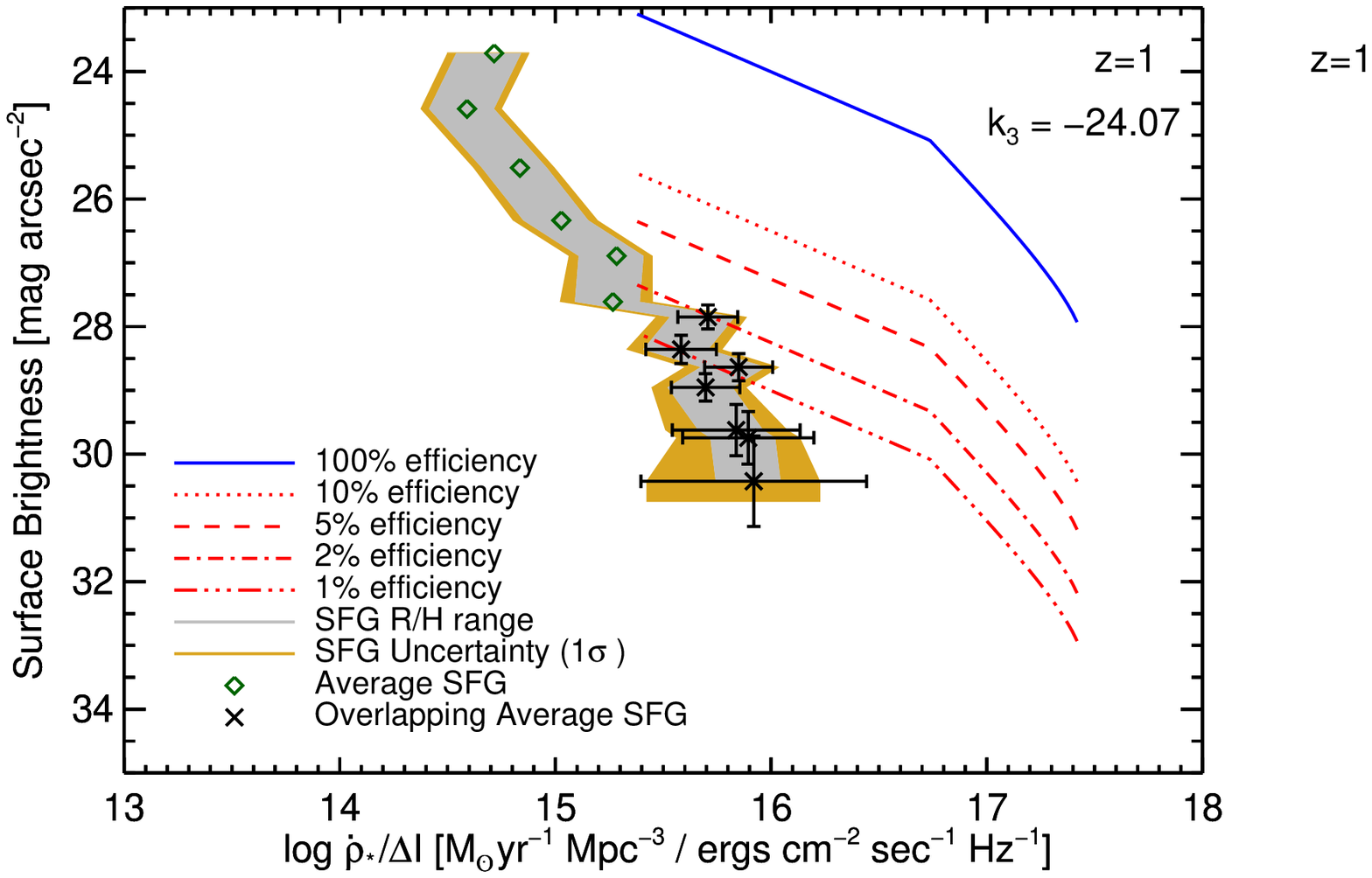}
\includegraphics[scale=0.4, viewport=10 5 490 340,clip]{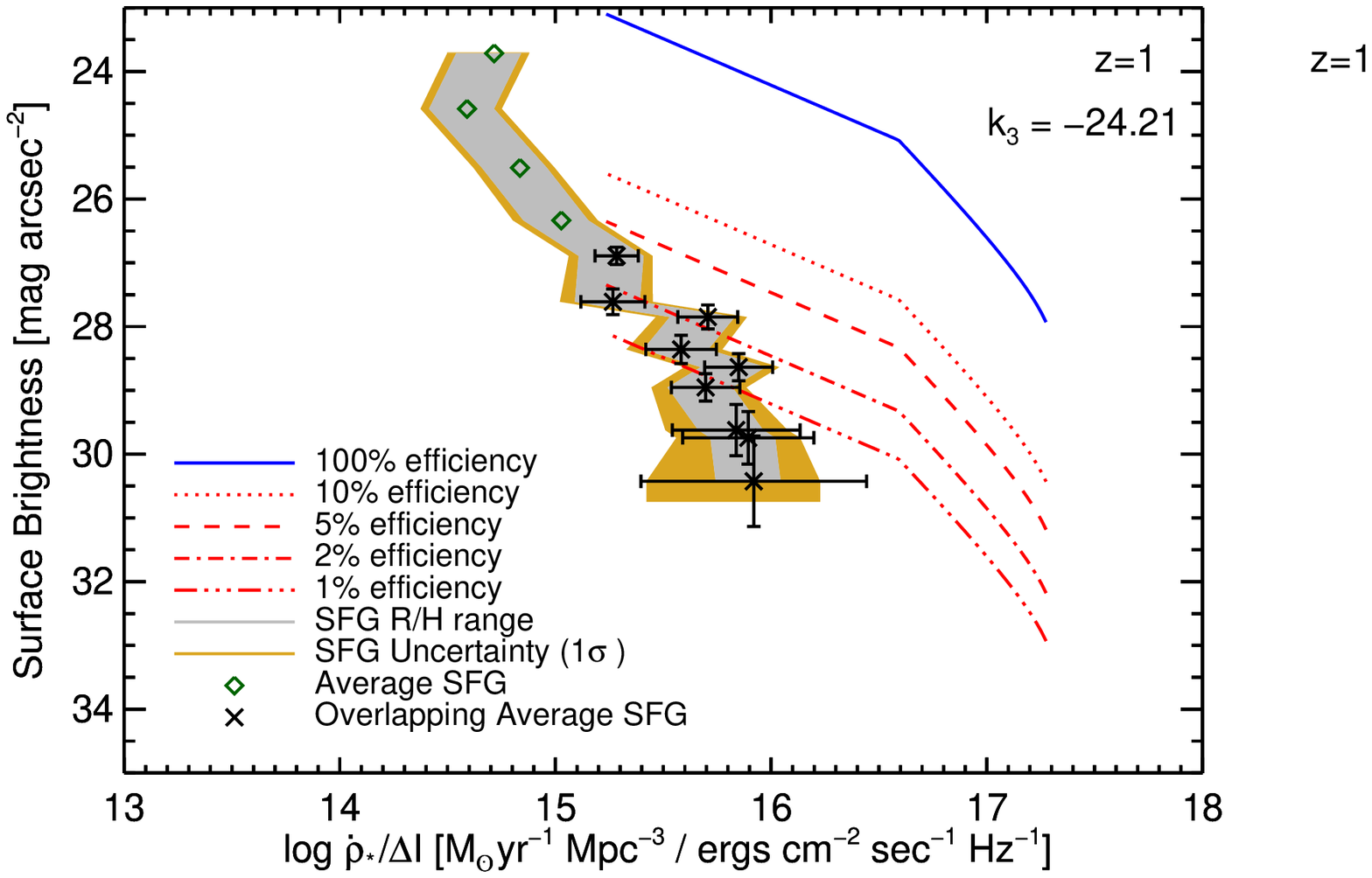}
\includegraphics[scale=0.4, viewport=10 5 490 340,clip]{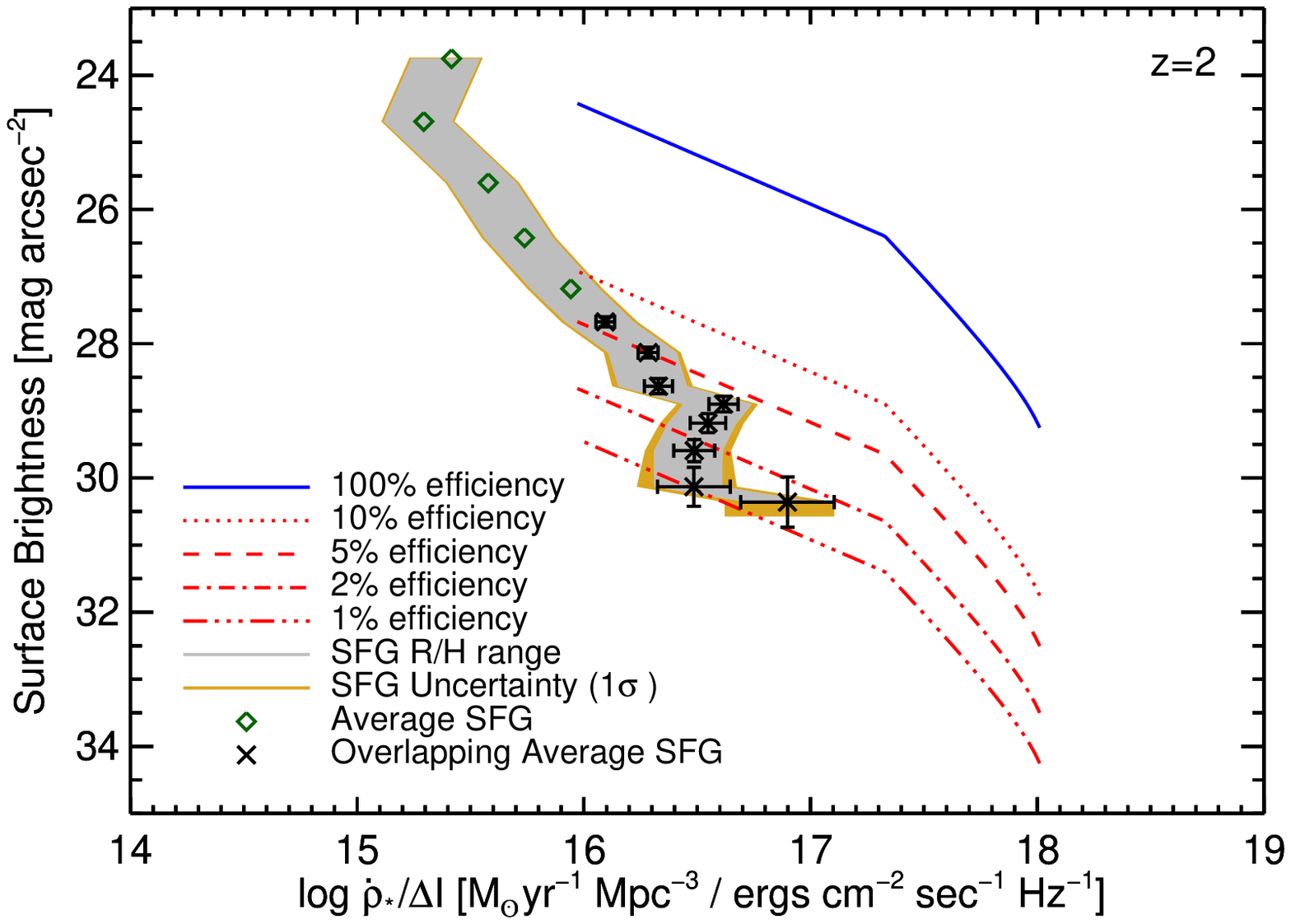}
\includegraphics[scale=0.4, viewport=10 5 490 340,clip]{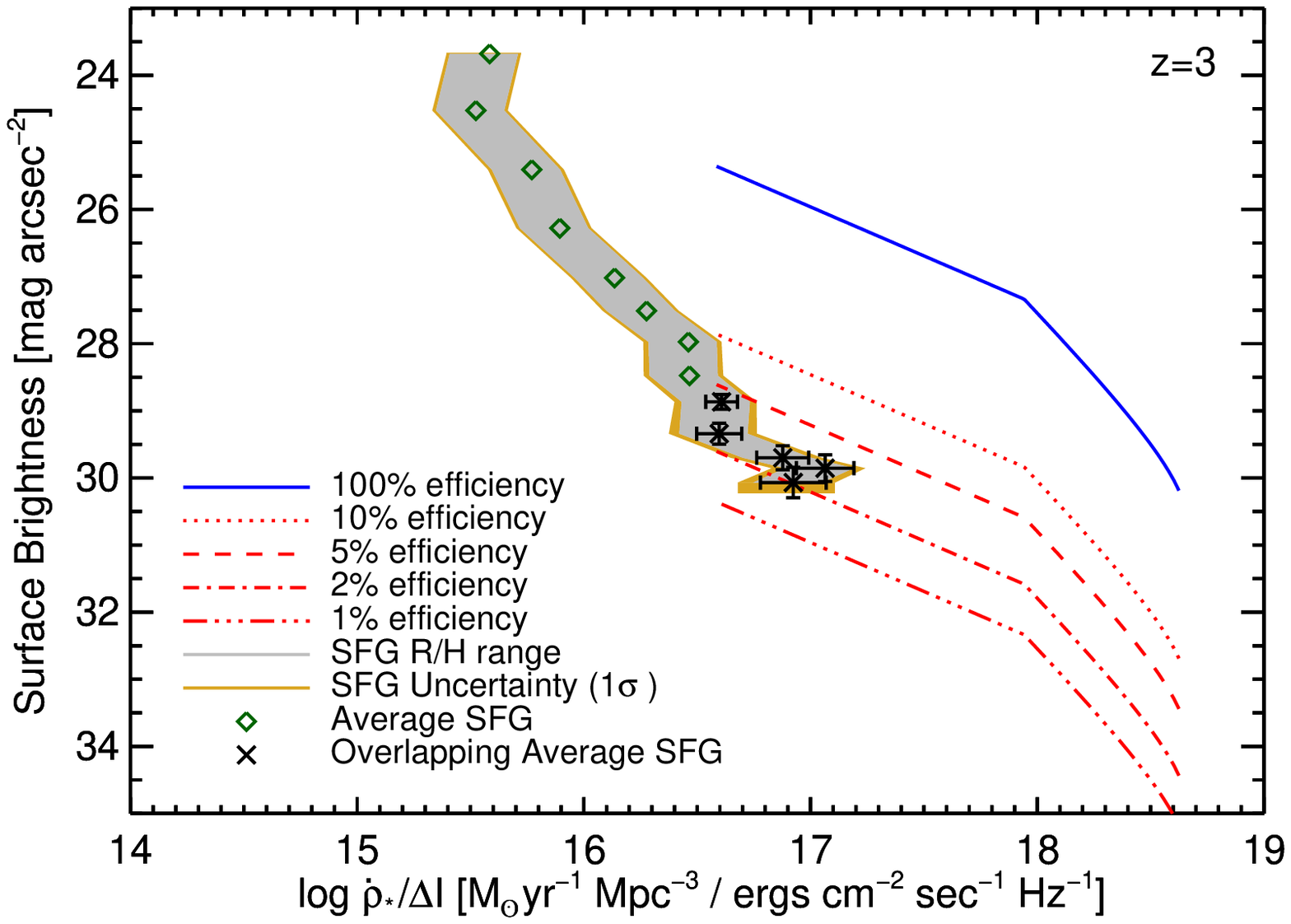}
}
\caption{ \label{fig:diffrhodotstar} 
Surface brightness versus the differential comoving SFR density per intensity 
($\Delta \dot{\rho_*}/\Delta\langle I_{\nu_0}^{\rm obs}\rangle$),  comparing the measured emission in the
outskirts of SFGs to the predicted levels for atomic-dominated gas for different SFR efficiencies. 
Each panel corresponds to a redshift bin, and there are two $z\sim1$ panels corresponding to the two possible
values of $k_3$. 
The blue line represents the model at 100\% efficiency, 
and the red dotted, short-dashed, dotted-dashed, and triple-dott-dashed lines represent 10\%, 5\%, 2\%, and 1\% efficiency. 
The filled gray region represents the observed emission in the outskirts of SFGs for a range in aspect ratios, and the filled gold region is its 1-$\sigma$ uncertainty.
The green diamonds are the average value of the possible aspect ratios,
with the error bars including the uncertainty due to the variance in the composite stack.
The SFR efficiency is obtained by comparing the measurements to the models, and show SFR efficiencies of 1-10\%.
}
\end{figure*}

\subsection{SFR efficiency determination}
\label{efficiency}
We determine the SFR efficiency of atomic-dominated neutral hydrogen gas at each redshift
by comparing the emission observed in the outskirts of SFGs in Figure \ref{fig:sbprofiles} to the emission expected from DLAs
based on a model which predicts the SFR density per intensity interval around the SFGs. This model, and the framework to connect
the observations to the model, are described in detail in Section 6 of \citet{Rafelski:2011}. 

The model adopts a disk-like geometry, the observed $f(N_{\rm HI})$, and the KS relation for different possible SFR efficiencies. 
The KS relation connects the star formation rate per unit area ($\Sigma_{\rm SFR}$) with the gas surface density ($\Sigma_{\rm gas}$), 
with parameters calibrated with observations of nearby disk galaxies \citep{Schmidt:1959, Kennicutt:1998a}.  The KS relation is defined as 

\begin{equation}
\Sigma_{\rm SFR} = K\times\left(\frac{\Sigma_{\rm gas}}{\Sigma_{\rm c}}\right)^\beta\;.
\label{eq:ksrelation}
\end{equation}

\noindent where $\Sigma_{\rm gas}$ is the mass surface density perpendicular to the plane of the disk, $\Sigma_{\rm c}=1$$M_\odot$pc$^{-2}$,
 $K=K_{\rm Kenn}$=(2.5$\pm$0.5){$\times$}10$^{-4}$ {\smpykpc}, and ${\beta}$=1.4$\pm$0.15 \citep{Kennicutt:1998a}. We define the SFR efficiency as
 the percentage change in the normalization $K$ in the KS relation. A 10\% SFR efficiency would correspond to a model which reduces the normalization 
 in the relation by a factor of 10, e.g. $K=2.5\times10^{-5}$ {\smpykpc}. We note that this is a different definition than what is often referred to as the 
 star formation efficiency \citep[e.g.][]{Bigiel:2010b}, although the two are directly related. 

In the next two subsections we guide the reader through the basic process to determine the SFR efficiency, but refer to \citet{Rafelski:2011} for details,
mathematical derivations, and equations. We describe the model in Section \ref{model}, the measurements and completeness corrections
in Section \ref{measure}, and compare the measurement and model to obtain the SFR efficiency in Section \ref{compare}. 

\subsubsection{Model for DLA emission}
\label{model}
The model by \citet{Rafelski:2011} is based on the model by \citet{Wolfe:2006}, and predicts the expected UV emission from DLAs by multiplying the expected SFR per area for each column density of gas from the KS relation \citep{Kennicutt:1998a} with the area of the sky covered by that gas from the column-density distribution function of H\textsc{i} gas \citep[e.g.][]{Prochaska:2009}.
The model assumes that SFGs are at the center of DLAs, and that
the in-situ star formation in DLAs occurs in gaseous disks inclined on the plane of the sky, averaged over all possible inclination angles. 
Each differential interval of the SFR density represents a ring around the SFGs corresponding to a surface brightness and a solid angle interval subtended by each ring. 
It allows for a large range in disk thicknesses, covering both very thick and thin disks (aspect ratios R/H from 10 to 100). The model is dependent on redshift by $(1+z)^3$ due to cosmological dimming and the K correction, and also depends on the redshift evolution of $k_3$ in $f(N_{\rm HI})$ shown in Figure \ref{fig:fofn_evol}.

The geometry of the model assumes that the gas column density of the disk (denoted as $g(N_{\perp},X)$ in \citet[][]{Rafelski:2011} and defined in equations 15 and 16 therein) decreases as a function of increasing radius, and directly depends on $f(N_{\rm HI},X)$. The $g(N_{\perp},X)$ also depends on the column density perpendicular to the disk, $N_{\perp}$, but since we integrate over all possible inclination angles, we do not need to know that quantity directly. By defining $f(N_{\rm HI},X)$ as a function of $g(N_{\perp},X)$ in this model, we implicitly assert that all the DLA gas is within a gaseous configuration with a preferred plane of symmetry and a decreasing column density as a function of radius, such as a disk. This results in a distribution of gas column densities that reproduce $f(N_{\rm HI},X)$, but with the gas in a disk-like structure described by $g(N_{\perp},X)$. We also consider the case where only half of the gas resides in such a disk (or only half contributes to the in-situ star formation) in Section 7.4 of \citet{Rafelski:2011} and also address it in Section \ref{models} below. 

We can predict the star formation rate density of the DLA gas as a function of the column density by applying the KS relation and integrating $g(N_{\perp},X)$ over all possible inclination angles of the disk. The comoving SFR density, $\dot{\rho_*}$, with observed column density greater than N, is derived in \citet{Wolfe:2006} and shown there as equation 6. This quantity is then modified in \citet{Rafelski:2011} to be a differential expression with respect to the the column density to model the change in $\dot{\rho_*}$ as a function of the column density  \citep[$\frac{d\dot{\rho_*}}{dN}$, equation 19 in][]{Rafelski:2011}, and therefore also the radius of the disk. In order to compare this quantity with observations, we replace the column density by the observed intensity averaged over all inclination angles. The model thereby predicts the surface brightness (obtained from the KS relation) as a function of the differential comoving SFR density per observed intensity interval, $d\dot{\rho_*}/{d\langle I_{\nu_0}^{\rm obs}\rangle}$ \citep[equation 25 in][]{Rafelski:2011}.

While this may seem to be a strange quantity, it enables a comparison of the model to the measured radial surface brightness profile, providing unique non-overlapping predictions
for possible SFR efficiencies. The SFR efficiency is modified in the model by varying the normalization of the KS relation. 
For the model prediction, we directly use equation 25 by \citet{Rafelski:2011}, along with the equations and constants it depends on therein. The only 
difference is that we modify the constants in $f(N_{\rm HI})$, adopting  the $k_3$ values for each redshift bin as described in Section \ref{fofn}.

The resulting predictions for the surface brightness versus $d\dot{\rho_*}/{d\langle I_{\nu_0}^{\rm obs}\rangle}$ for each redshift window are shown in Figure \ref{fig:diffrhodotstar}, 
where the blue line represents 100\% SFR efficiency, and the red lines correspond to reduced efficiencies as labeled in the Figure. A 10\% SFR efficiency is obtained by reducing the 
normalization constant K by a factor of 10, where 100\% efficiency corresponds to $K_{\rm Kenn}$=(2.5$\pm$0.5){$\times$}10$^{-4}$ {\smpykpc} \citep{Kennicutt:1998a}.
Since there are two possible values of $k_3$ found at $z\sim1$, we show the $z\sim1$ results for two different $k_3$ values.

\subsubsection{Measurement and completeness correction}
\label{measure}
In order to obtain a SFR efficiency below, we need to compare the model to the observations, which requires putting the observations into the same form as the model. Specifically, the radial surface brightness profiles from Figure \ref{fig:sbprofiles} showing emission observed in the outskirts of SFGs need to be transformed onto the same surface brightness versus $d\dot{\rho_*}/{d\langle I_{\nu_0}^{\rm obs}\rangle}$ scale as the model. 
For each ring around the stacked SFGs, corresponding to a point in the radial surface brightness profile, we obtain a surface brightness and covering area and use it to determine $\Delta \dot{\rho_*}$ with equation 28 by \citet{Rafelski:2011}. This equation depends on the number of SFGs in each redshift bin (from the full sample), the angular diameter distance, and the 
comoving volume sampled, which is all provided in Table \ref{tab:samp}. It also depends on the aspect ratio of the disk, which is an unknown quantity, but is likely contained in the range between 10 to 100, and therefore the resultant measurements of $\Delta \dot{\rho_*}/{\Delta \langle I_{\nu_0}^{\rm obs}\rangle}$ have a range of possible values.

We obtain ${\Delta \langle I_{\nu_0}^{\rm obs}\rangle}$ by measuring the intensity change across each ring by taking the difference of the intensity on either side of each point and dividing by two. If ${\Delta \langle I_{\nu_0}^{\rm obs}\rangle}$ were negative, then we would increase the interval over which we measure ${\Delta \langle I_{\nu_0}^{\rm obs}\rangle}$, as this would be due to noise fluctuations. This mainly occurs at radii that are later excluded due to low signal-to-noise. The resultant determination of $\Delta \dot{\rho_*}/{\Delta \langle I_{\nu_0}^{\rm obs}\rangle}$ is therefore obtained from the combination of the surface brightness, covering area, and intensity decrease over each ring. 
The redshift dependence of $\Delta \dot{\rho_*}/{\Delta \langle I_{\nu_0}^{\rm obs}\rangle}$ includes the same cosmological dimming and K correction assumed in the model, and also includes the redshift in the angular diameter distance and in the comoving volume provided in Table \ref{tab:samp}. 

Before adding the observations to Figure \ref{fig:diffrhodotstar}, we have to consider the completeness of the observations. First, we applied a magnitude cut of $V<29$ to the full sample to ensure
robust photometric redshifts. However, galaxies fainter than this magnitude cut, and galaxies below the detection threshold, 
would presumably also be surrounded by DLA gas, and thus some of the star formation will be missed. The details of the completeness corrections are provided in Appendix A3 by \citet{Rafelski:2011}, 
and we outline the procedure here. We first determine the number of missed galaxies from extrapolations of the luminosity functions of SFGs for each redshift bin integrated out to $V\sim33$, where most of the contribution comes from galaxies at $29<V<31$.
We use the luminosity function from \citet{Oesch:2010eb} for $z\sim1$, \citet{Alavi:2014} for $z\sim2$, and \citet{Reddy:2009} for $z\sim3$, applying k corrections of 0.24, 0.17, and 0.15 respectively. 

In this correction, we assume that the shape of the radial surface brightness profile does not change for the fainter galaxies, and then for each ring 
we consider the number of SFGs missed, and the fraction of flux missed for these galaxies in half magnitude bins by scaling the profile to the integrated flux of the missed galaxies.
This yields the amount of $\Delta \dot{\rho_*}/{\Delta \langle I_{\nu_0}^{\rm obs}\rangle}$
missed for each ring, which is added to the measurement of $\Delta \dot{\rho_*}/{\Delta \langle I_{\nu_0}^{\rm obs}\rangle}$.
This results in an increase in $\Delta \dot{\rho_*}/{\Delta \langle I_{\nu_0}^{\rm obs}\rangle}$ by 60-80\% depending on the redshift, and 
the magnitude of the completeness correction can be visualized by comparing figure 8 and figure 9 in \citet{Rafelski:2011}. 

In Figure \ref{fig:diffrhodotstar}, we plot the completeness corrected data using the same conventions as \citet{Rafelski:2011}, where the green diamonds and black crosses are for average values of the aspect ratio. 
The change of color and symbols marks the transition from UV emission that could be from atomic-dominated gas (black) to the inner regions of the SFGs that
comes from molecular dominated gas (green), which is discussed further in Section \ref{compare}. The error bars include the uncertainty in the aspect ratios, 
the variance due to stacking different SFGs obtained from the bootstrap method described above, and the measurement uncertainties.
The gray shaded regions represents the range in aspect ratios, and the gold shaded regions the uncertainty on that range.  We do not include possible systematic uncertainties 
in the FUV to SFR conversion factor, the column-density distribution function, or any such systematics. 
The data shown in Figure \ref{fig:diffrhodotstar} are truncated at a maximum radius corresponding to a 3-$\sigma$ cut in Figure \ref{fig:sbprofiles}.

\subsubsection{Comparison of model and measurement}
\label{compare}

With the differential $\dot{\rho_*}$ as a function of the radius from the central SFGs for both the model and the data, we can equate the two to determine a SFR efficiency. Doing so assumes that the FUV emission in the outskirts of the SFGs is on average emitted from the in-situ star formation from atomic-dominated gas contained in a disk-like structure around these galaxies. As described in the introduction, the association of the DLAs with these SFGs is well established, and in addition this comparison assumes that the gas is in a disk-like geometry surrounding the SFGs. This assumption requires that the covering fraction of these disk-like structures around the SFGs approximately matches the covering fraction of the gas, which is checked in Section \ref{cf}. In addition, there must be some point at which the molecular-dominated central star-forming cores transition to the atomic-dominated disk-like structure imposed by this comparison. 

The transition from what is likely atomic-dominated to molecular-dominated gas in the data is determined by comparing the data and the model to each other, and finding
the value of $d\dot{\rho_*}/{d\langle I_{\nu_0}^{\rm obs}\rangle}$ in the model matching the measurement. 
We truncate the model at $N_{\rm HI}=$1.2$\times$10$^{22}$ cm$^{-2}$, as above these column densities DLAs are not frequently observed, likely due to the conversion from atomic to molecular hydrogen gas \citep{Schaye:2001b}. This cut is marked by the left side of the red/blue model lines and by the color change in the data points from black to green. A small number of DLAs exist at higher column densities \citep{Noterdaeme:2014,Noterdaeme:2015}, but this value is more typical and consistent with \citet{Rafelski:2011}.

The exact choice of the maximum $N_{\rm HI}$ does not affect the resultant SFR efficiencies, just the point
at which we consider the gas to be transitioning from atomic-dominated gas to molecular-dominated gas, and therefore the transition from green to black points in Figure \ref{fig:diffrhodotstar} \citep{Krumholz:2009c,Krumholz:2009a}.
We acknowledge that there is likely a transition region where the gas is not fully in a single phase, and thus it's possible that our most luminous atomic-dominated point may not be from
purely atomic-dominated star formation.  

The models and the data do appear to have somewhat different slopes, which suggests that there may be a change in the efficiency of the gas as a function of the radius of the SFGs. This effect appears to be stronger in the lower redshift data than that at $z\sim3$, although the uncertainties on the data that suggest a different slope are also larger. While this change in efficiency as a function of radius may be real, we believe the data are not sufficiently good to make accurate measurements of the slope, and therefore this effect. If real, it would suggest that the efficiency is decreasing as we go out in radius, but better data would be required to investigate the physical origin of this trend.

With the model and observations on the same plot, it is straightforward to obtain the SFR efficiency of each ring around the SFGs by comparing the model and the measurements in the atomic-dominated regime.
For each measured point corresponding to a ring around the SFGs, we match it to the overlapping model (with a finer grid than plotted), and thereby obtain the SFR efficiency from that model. 
Overall, the SFR efficiency of neutral atomic-dominated hydrogen gas ranges from $\sim$1-6\% depending on the surface brightness (and therefore $N_{\rm HI}$), with mean values around 2\% as shown in Table \ref{tab:SFR}.

\begin{deluxetable}{ccc}[h!]
\tablecaption{Average SFR Efficiency \label{tab:SFR}}
\tablehead{
\colhead{Redshift} &
\colhead{log $k_3$} &
\colhead{Efficiency}
}
\startdata
$z\sim1$ & -24.26 & $1.4\pm 1.0$   \\
$z\sim1$ & -24.07 & $0.8\pm 0.6$  \\
$z\sim2$ & -24.00 & $3.1\pm1.7$  \\
$z\sim3$ & -23.77 & $2.8\pm 0.7$
\enddata
\tablecomments{Average SFR efficiency (based on the normalization of the KS relation, see Section \ref{efficiency}) for different redshifts and log $k_3$. The uncertainty is the standard deviation of the various efficiencies measured for different surface brightnesses (and therefore $N_{\rm HI}$), and not the measurement uncertainty. }
\end{deluxetable}

\subsection{Visualizing the SFR efficiency}
\label{vis}

At this point we have determined a SFR efficiency for each ring in the radial surface brightness profile, but this quantity is difficult to compare between redshifts, and more importantly, with other studies. 
As a tool to understand the low SFR efficiencies, to compare the efficiencies in different redshift bins to each other, and to compare the results to other studies, we translate the results to a common set of quantities, namely
$\Sigma_{\rm SFR}$ and $\Sigma_{\rm gas}$.
 We emphasize that at this point the SFR efficiency is already measured, and that this is purely a visualization tool. 
The details of this conversion is described in Section 6.3 in \citet{Rafelski:2011}, but we guide the reader through the process here.

\begin{figure*}[h!]
\center{
\includegraphics[scale=0.4, viewport=10 5 490 350,clip]{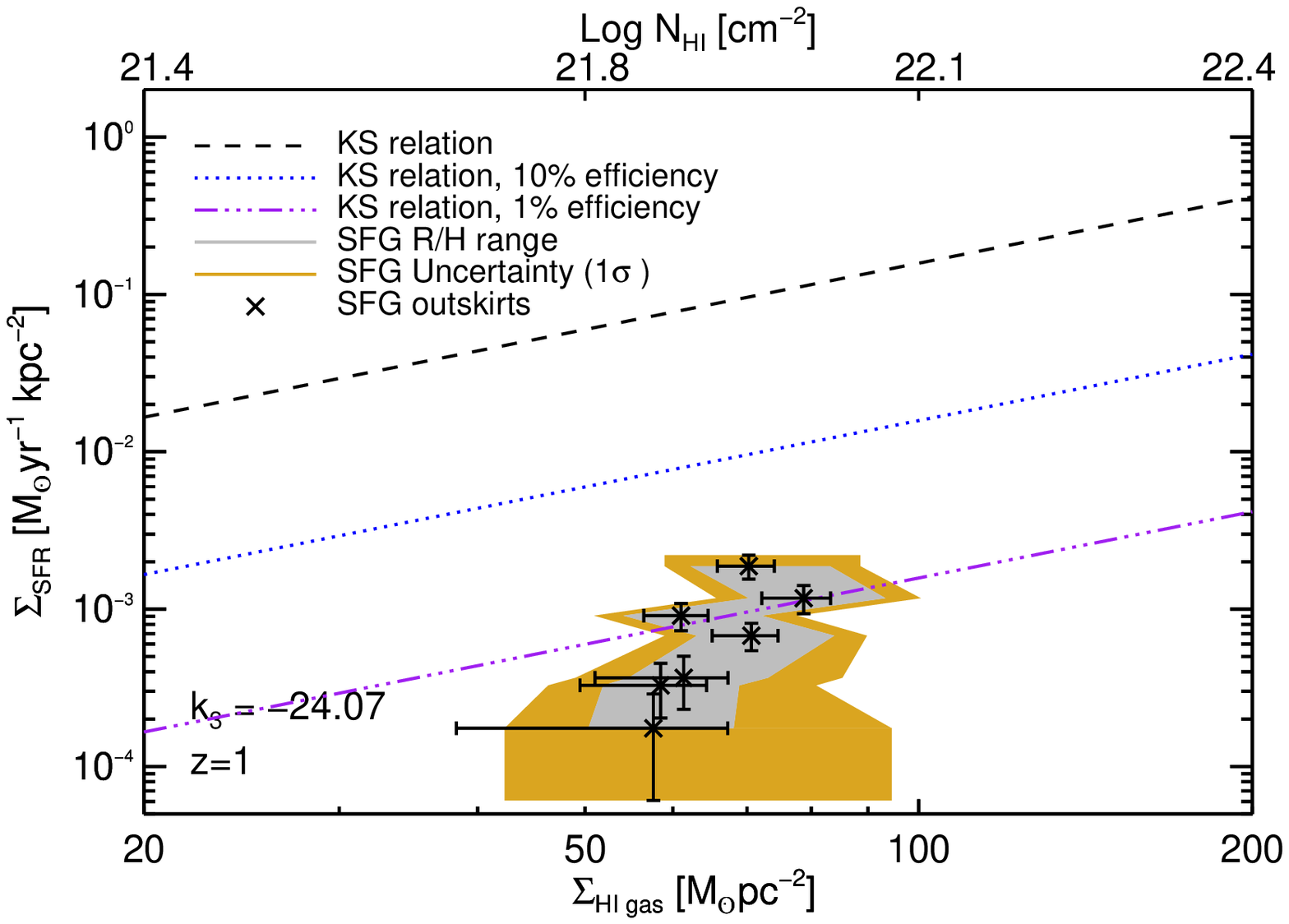}
\includegraphics[scale=0.4, viewport=10 5 490 350,clip]{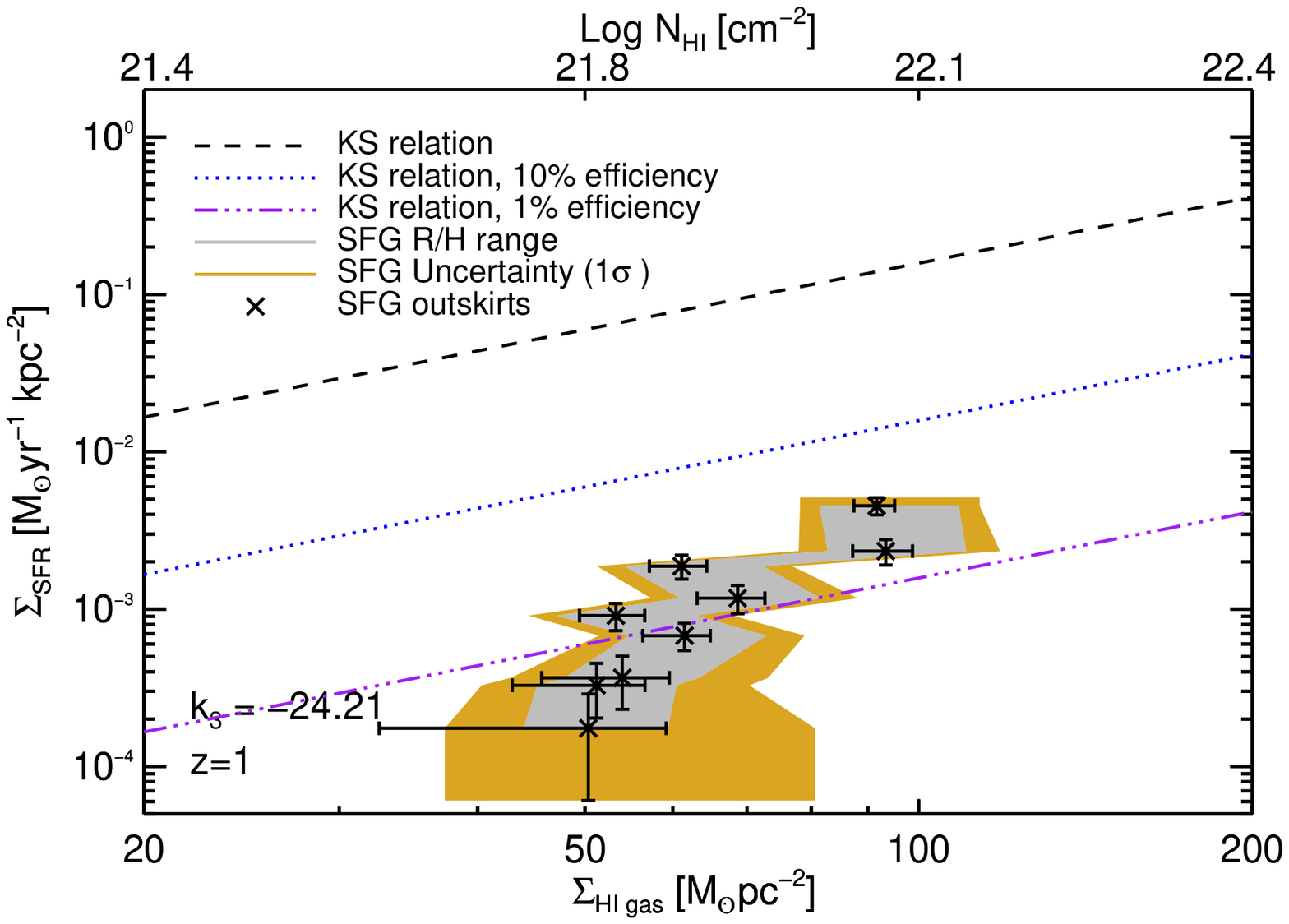}
\includegraphics[scale=0.4, viewport=10 5 490 350,clip]{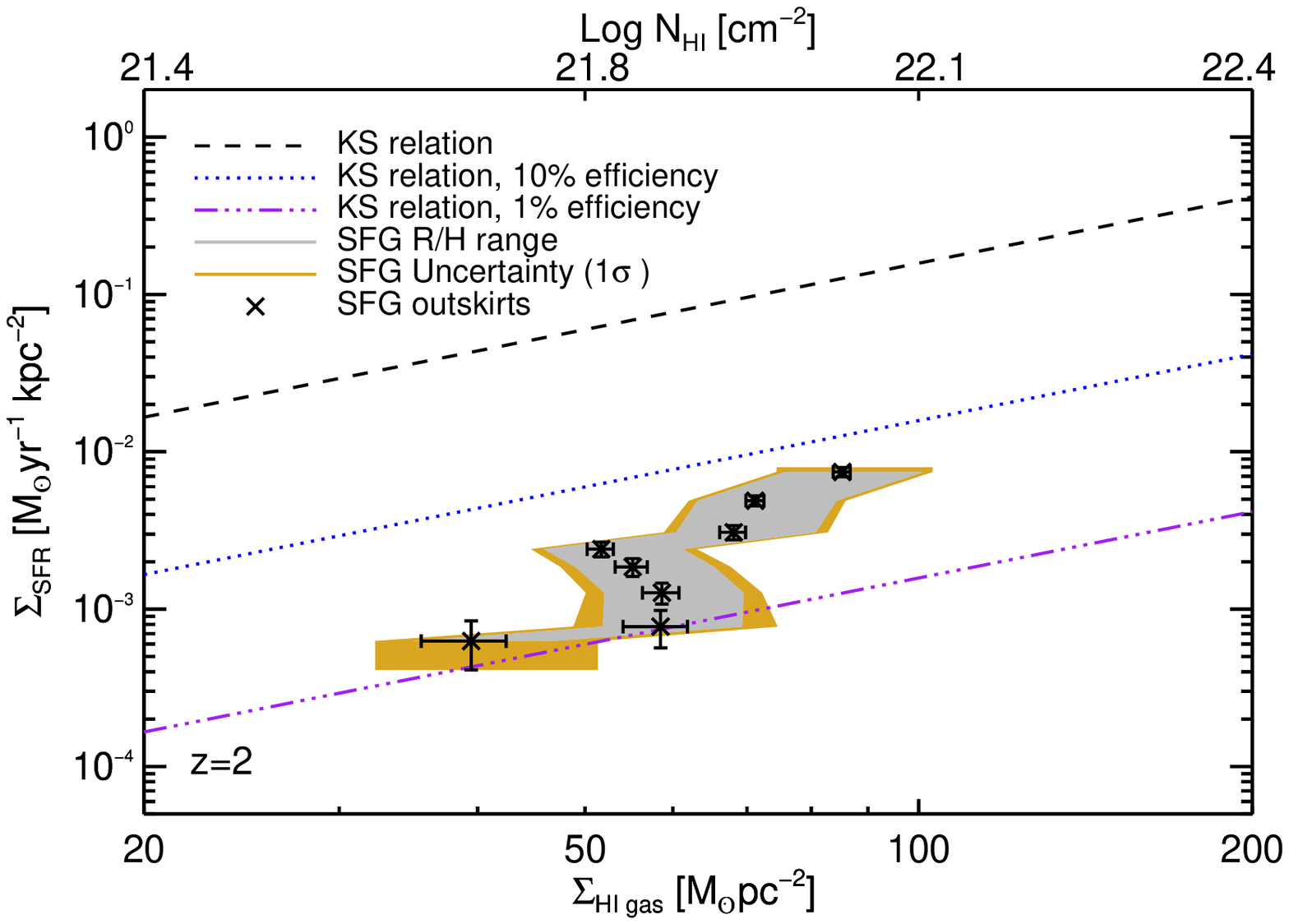}
\includegraphics[scale=0.4, viewport=10 5 490 350,clip]{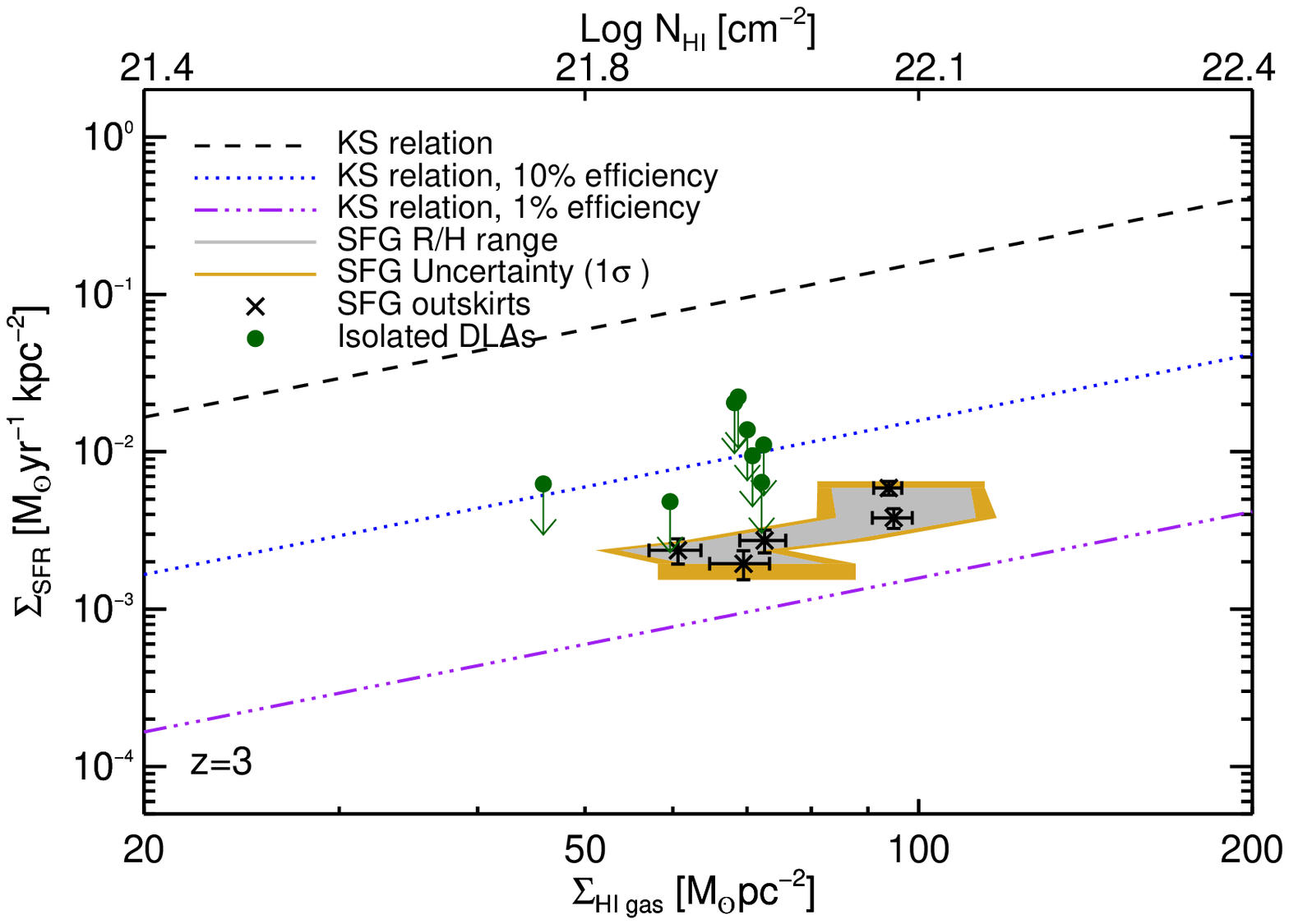}
}
\caption{ \label{fig:sfr_sigma} 
Star formation rate per unit area ($\Sigma_{\rm SFR}$) versus gas density ($\Sigma_{\rm gas}$) used to visualize the reduced SFR efficiency. 
Each panel corresponds to a redshift bin, and there are two $z\sim1$ panels corresponding to the two possible values of $k_3$. 
The dashed black line represents the KS relation for 100\% SFR efficiency, the dotted blue line is for 10\% efficiency, and the dotted-dashed purple
line is for 1\% efficiency. 
The gray filled region, the gold filled region, and the black crosses represent the same data as in Figure \ref{fig:diffrhodotstar}. 
The green data points in the $z\sim3$ panel correspond to upper limits derived
for DLAs without central bulges of star formation from \citet{Wolfe:2006} converted to work with this plot by \citet{Rafelski:2011}.
The data all fall below 10\% of the KS relation at all redshifts, showing a reduced SFR efficiency. However, no clear evolution with redshift in the SFR efficiency is observed between $z\sim1$ to $z\sim3$. }
\end{figure*}

The $\Sigma_{\rm SFR}$ can be directly obtained from the intensity of the emission in the outskirts of the SFGs as measured in the radial surface brightness profile by equation 4 in \citet{Rafelski:2011}.
This averages the SFGs over all disk inclination angles and applies the KS relation and the FUV to SFR conversion to obtain $\Sigma_{\rm SFR}$ directly from the intensity. 
The corresponding $\Sigma_{\rm gas}$ is indirectly obtained by taking $\Sigma_{\rm SFR}$ and plugging it into the the KS relation for the measured reduced SFR efficiency. 
We plot $\Sigma_{\rm SFR}$ versus $\Sigma_{\rm gas}$ for each redshift bin in Figure \ref{fig:sfr_sigma}. We note that the corresponding column density for $\Sigma_{\rm gas}$ is obtained by unit conversion and the number of atoms per solar mass. 

\begin{figure*}[]
\center{
\includegraphics[scale=0.6, viewport=10 5 490 352,clip]{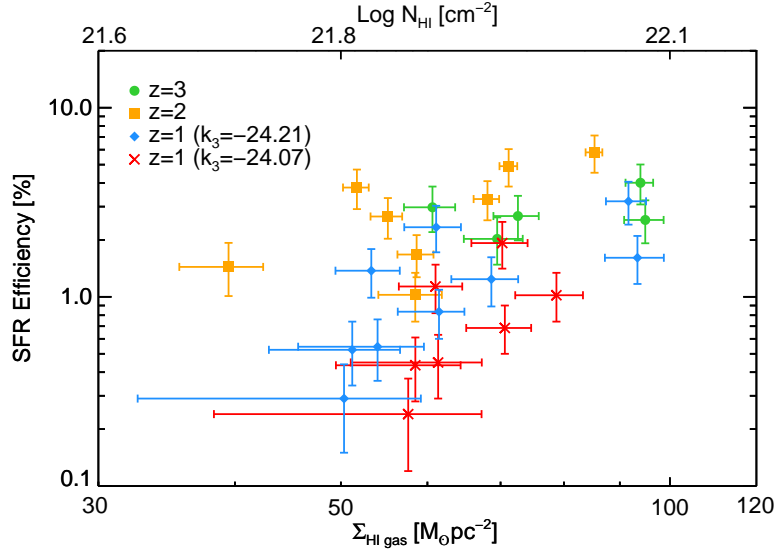}
}
\caption{ \label{fig:efficiency} 
SFR efficiency versus implied $\Sigma_{\rm gas}$ shown for different redshifts. The green circles correspond to the $z\sim3$ SFGs, the orange squares to those at $z\sim2$, and the blue diamonds and red crosses are for those at $z\sim1$. The blue diamonds correspond to log $k_3=-24.26$, obtained by a fit to the $z>2$ $k_3$ values. The red crosses are for log $k_3=-24.07$, obtained by an interpolation of the $z\sim2$ point and the poor fit to the $z\sim0$ data. All the data except the red $z\sim1$ points show no evolution with redshift, while the red points may suggest a mild decrease at lower redshift. However, see the discussion in Section \ref{discussion}.}
\end{figure*}

We make a more direct comparison of the SFR efficiencies as a function of redshift in Figure \ref{fig:efficiency}, showing the SFR efficiency as a function of $\Sigma_{\rm gas}$ for each redshift bin. 
The variation in the $z\sim1$ points show the importance of the value of $k_3$ from the column-density distribution function in determining the SFR efficiency. The blue points in the figure correspond to the linear fit of the $k_3$ values in Figure \ref{fig:fofn_evol}, while the red points correspond to the extrapolation to the poor $z\sim0$ fit. The points show little evidence of evolution with redshift, except potentially the $z\sim1$ red points, but see the discussion in Section \ref{k3}. We emphasize that by construction our results average the data on the scale of a few hundred parsecs, and thus we measure an efficiency that is not purely the local one. In addition, the observations are only sensitive to the highest column-density DLAs, resulting in measurements of $\Sigma_{\rm gas}$ at the high end of $f(N_{\rm HI})$. However, based on the radius at which we measure the emission, 
we expect to measure the highest $N_{\rm HI}$ DLAs based on the correlation of $N_{\rm HI}$ with impact parameter $N_{\rm HI}$ \citep{Peroux:2016}.

\section{Covering Fraction}
\label{cf}

Throughout this paper we assume that the outskirts of SFGs form stars in-situ out of neutral atomic-dominated hydrogen gas. 
In order for this assumption to hold, the area on the sky covered by the emission around SFGs must be similar to the area covered by this type of gas at that redshift. 
At the same time, we could ask the reverse question; wether or not the area on the sky covered by molecular-dominated gas can account for this emission. 
To address this, we compare the covering fraction of the emission to that of atomic-dominated gas in Section \ref{atomic} and to molecular-dominated gas in Section \ref{molecular}, 
following the same methodology as described in Section 5.2 of \citet{Rafelski:2011} and summarized here. 

The cumulative covering fraction, $C_A$, is obtained by integrating the hydrogen column-density distribution function $f(N_{\rm H},X)$
for gas columns greater than column density $N$ up to a maximum column $N_{\rm max}$. This holds for both atomic and molecular hydrogen gas, 
with 

\begin{equation}
C_A(N) = {\int_{X_{\rm min}}^{X_{\rm max}}}dX {\int_{N}^{N_{\rm max}}dN_{\rm H} f(N_{\rm H},X)}\;,
\label{eq:covfrac}
\end{equation}

\noindent where $X$ is the absorption distance. $dX$ is defined as

\begin{equation}
dX \equiv \frac{H_0}{H(z)}(1+z)^2dz\;,
\label{eq:X}
\end{equation}

\noindent where $H(z)$ is the Hubble parameter at redshift $z$, and $H_0$ is the Hubble constant.
The column-density distribution function
is different for atomic-dominated and molecular-dominated hydrogen gas, and we consider the
covering fraction for each below.

\begin{figure*}[]
\center{
\includegraphics[scale=0.4, viewport=10 5 490 350,clip]{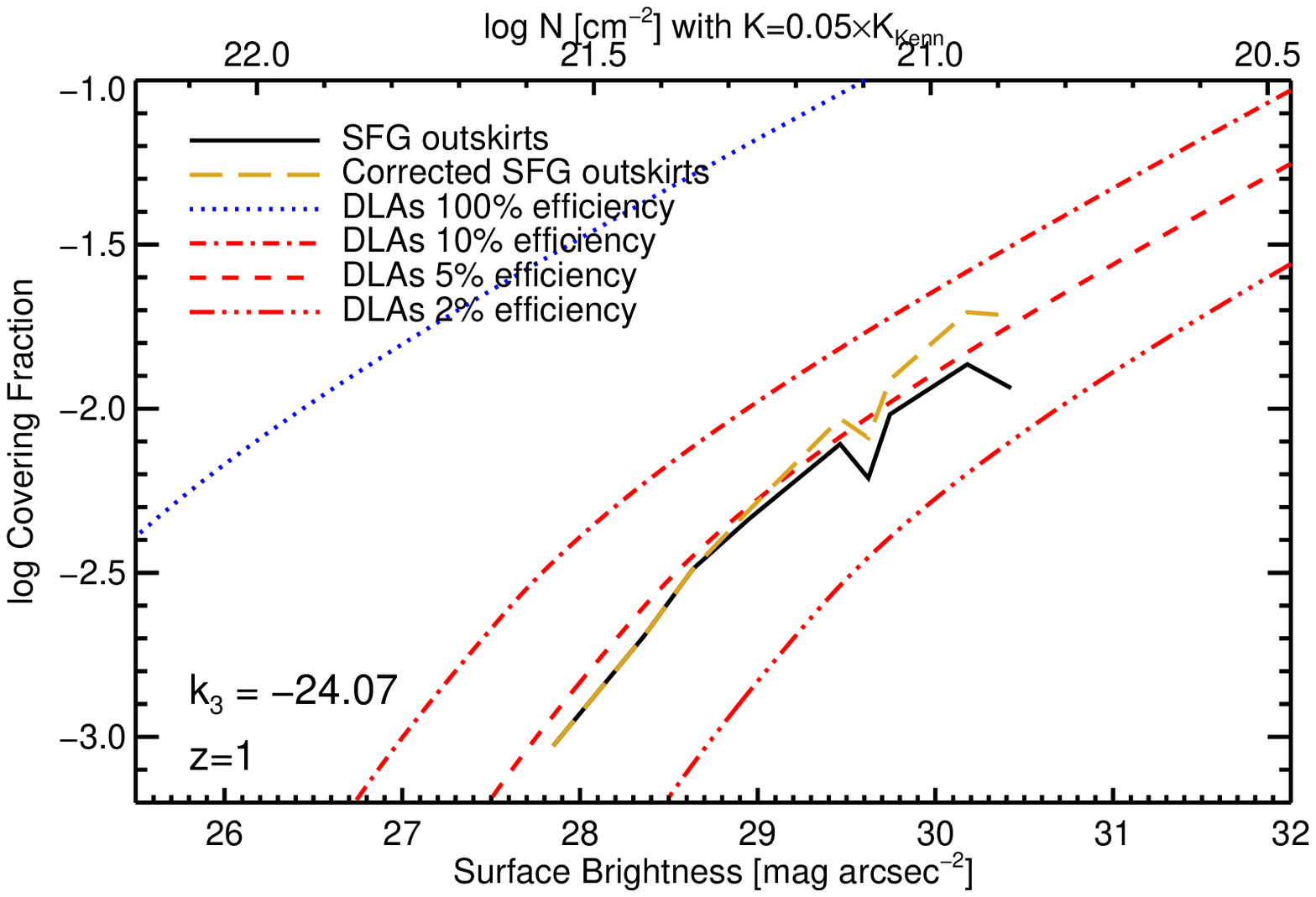}
\includegraphics[scale=0.4, viewport=10 5 490 350,clip]{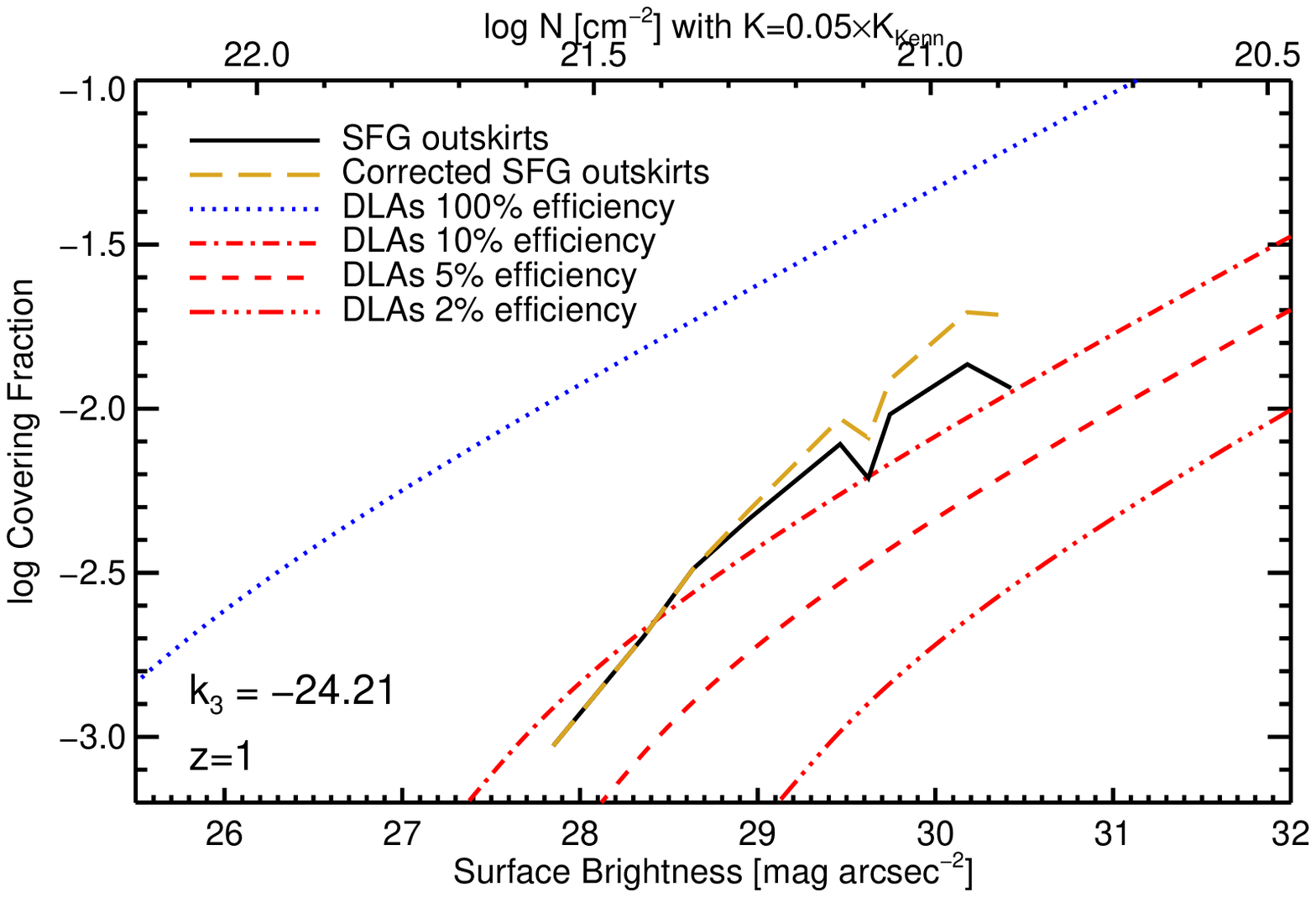}
\includegraphics[scale=0.4, viewport=10 5 490 350,clip]{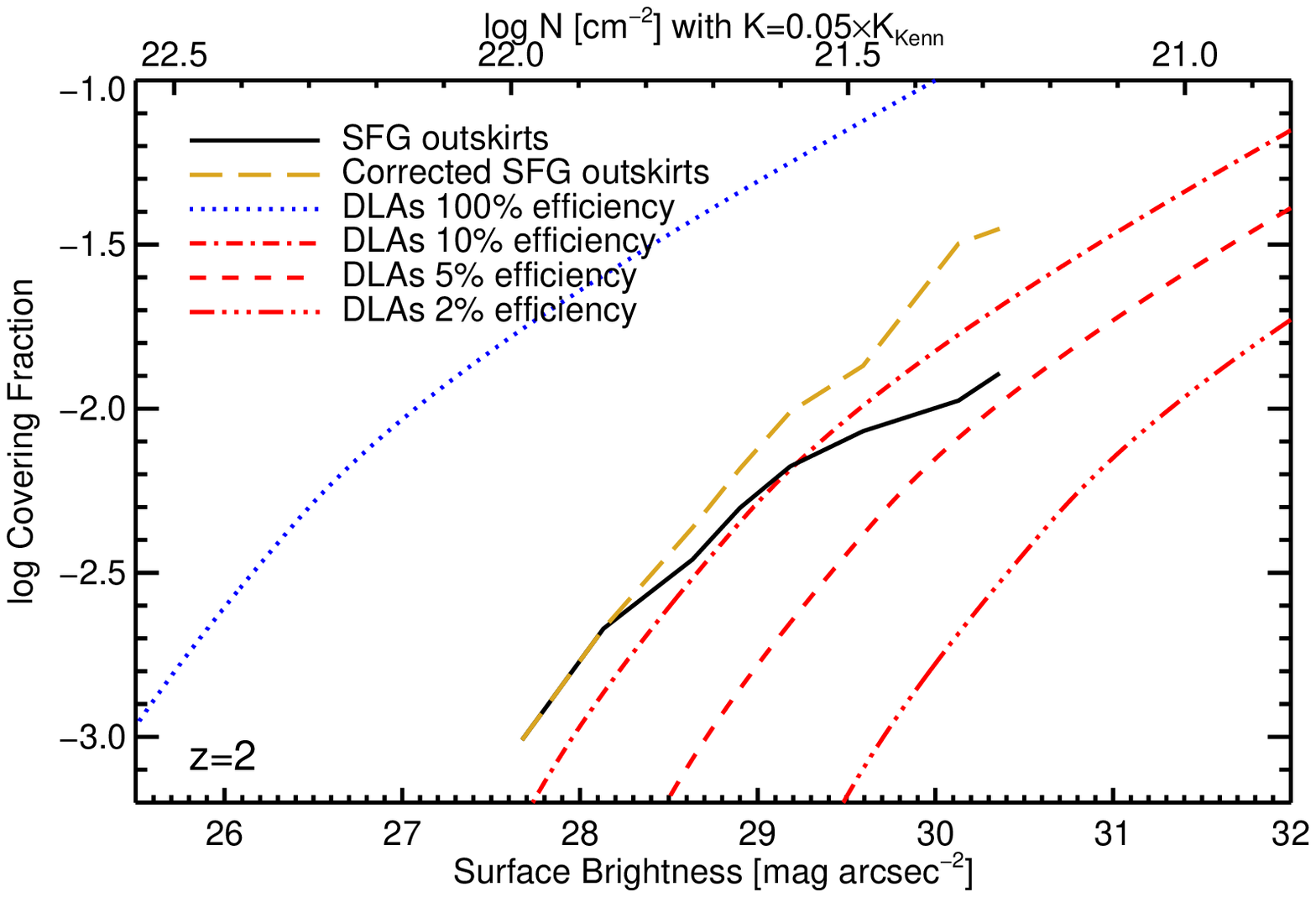}
\includegraphics[scale=0.4, viewport=10 5 490 350,clip]{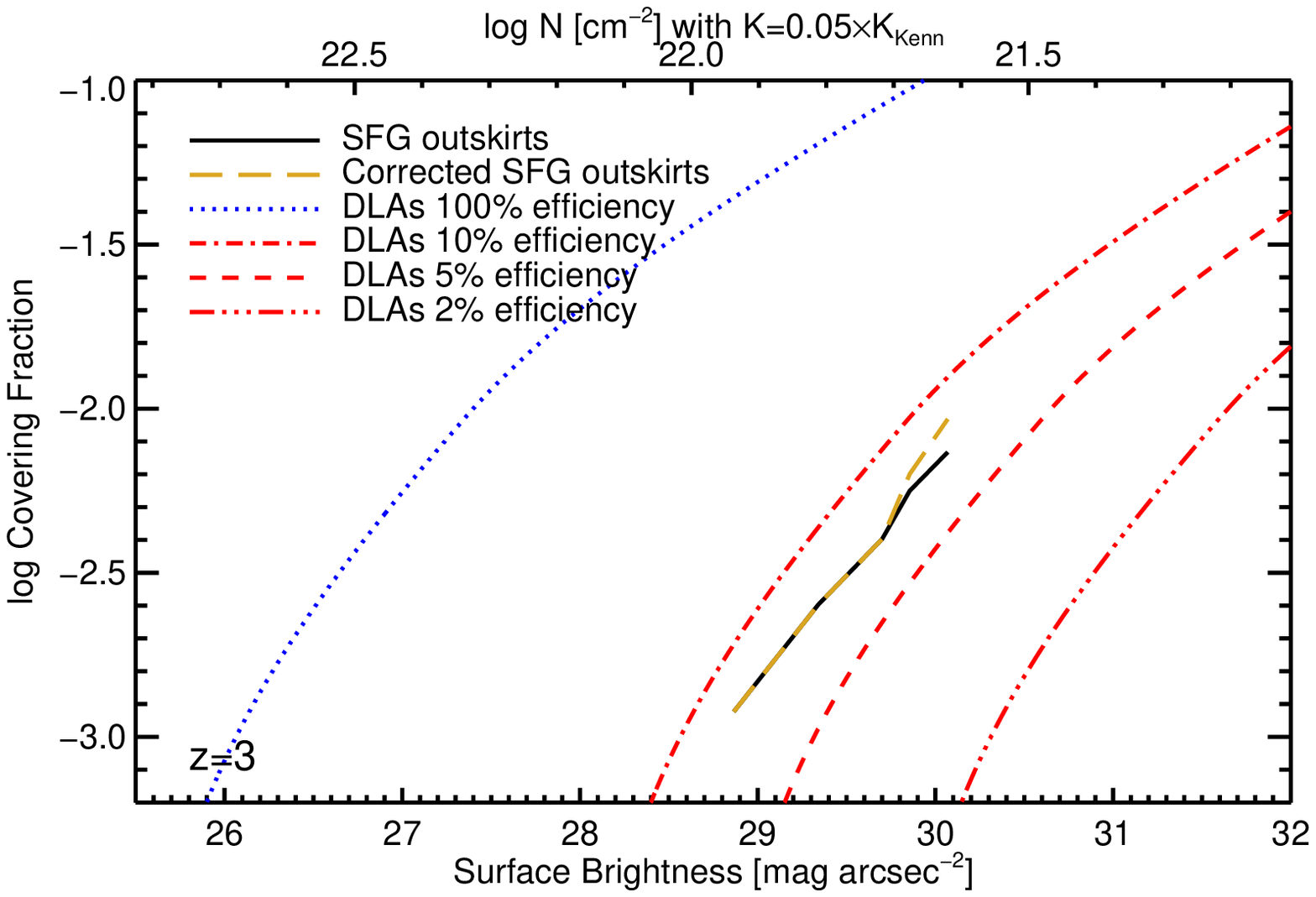}
}
\caption{ \label{fig:covfracHI} 
Cumulative covering fraction of DLAs as a function of surface brightness compared to the covering fraction of the outskirts of SFGs. 
The surface brightness depends both on the column density of the DLAs and the SFR efficiency. 
The black line is the covering fraction of the SFG outskirts, and the gold dashed line is the same corrected for completeness. 
The blue dotted line is the covering fraction for DLAs forming stars according to the KS relation at 100\% efficiency, 
and the red dotted-dashed, short-dashed, and triple-dott-dashed lines represent 10\%, 5\%, and 2\% efficiency. 
The column densities labeled at the top of each panel are for a 5\% efficiency, similar to the measurements. 
The general agreement of the covering fraction at similar SFR efficiencies show that there is suficient DLA gas to account for the emission observed in the outskirts of SFGs. 
}
\end{figure*}

\subsection{Covering Fraction of Atomic-dominated Gas}
\label{atomic}

We investigate whether the covering fraction of neutral atomic-dominated hydrogen gas (i.e. DLAs) is consistent with the covering fraction 
of the outskirts of SFGs. To do so we integrate Equation \ref{eq:covfrac} using the best fit parameters for $f(N_{{\rm H  I}},X)$
described in Section \ref{fofn}. Since we have two possible values of $k_3$ at $z\sim1$, we have two covering fraction possibilities for that redshift. 
We use the same $N_{\rm max}$ of $N_{\rm HI}=$1.2$\times$10$^{22}$ cm$^{-2}$ as in Section \ref{compare}. We note that the observables used for the 
covering fraction comparison are the same as used in the efficiency measurement, and therefore by construction this is not a completely independent 
measurement. However, this comparison is less model dependent than the efficiency measurement and is thus an important cross-check.

We compare the cumulative covering fraction of DLAs to that of the SFG outskirts as a function of surface brightness (and therefore column density) in Figure \ref{fig:covfracHI}.
The blue dotted line is the covering fraction for DLAs forming stars according to the KS relation at 100\% efficiency, 
and the red dotted-dashed, short-dashed, and triple-dott-dashed lines represent 10\%, 5\%, and 2\% efficiency. 
The efficiency is again reduced by 
reducing the normalization of the KS relation. The black line is the covering fraction of the SFG outskirts, which is obtained by the total 
area covered by the radial surface brightness profile above that surface brightness times the number number of SFGs in the full sample for that redshift bin,
divided by the total area of 11.40 arcmin$^2$. Since this number depends on the number of observed galaxies, it requires a completeness 
correction for fainter objects not included in the full sample. 

The completeness correction is obtained following Equation A1 from Appendix A1 in \citet{Rafelski:2011}. We first determine the number of missed galaxies from the 
luminosity functions described in Section \ref{measure} for each redshift bin. For each half-magnitude bin of missed galaxies, we scale the radial surface brightness profile to that magnitude,
and determine the area that the outskirts of the scaled profile cover for each surface brightness. The missed covering fraction is the sum over the half-magnitude bins
of the missed area times the missed number of galaxies. The resulting completeness corrected covering fraction as a function of surface brightness is shown in gold in Figure \ref{fig:covfracHI}.
While the number of galaxies increases quickly with decreasing galaxy luminosity, the scaling of the outskirts of the radial surface brightness profile drops more rapidly, and thus only the bright end of 
the luminosity function contributes. We note that we only consider the outskirts of the SFGs, which leads to a different correction than would be needed for molecular gas in Section \ref{molecular}.

A comparison of the completeness corrected covering fraction of the SFG outskirts and DLAs in Figure \ref{fig:covfracHI} reveal that the two agree at a somewhat higher SFR efficiency than expected (e.g. Figure \ref{fig:efficiency}). Specifically, the covering fractions agree for SFR efficiencies of $\sim$5-20\%, compared to the measured efficiencies of $\sim$1-5\%, resulting in differences by factors of $\sim$3 in efficiency, or $\lesssim5$ in covering fraction. This discrepancy in the SFR efficiency is not a large concern, as there are many assumptions that could cause systematic uncertainties on both the covering fraction and SFR efficiency. For instance, the covering fraction is directly dependent on the transition point from molecular-dominated to atomic-dominated gas, marked by the change from green to black points in Figure \ref{fig:diffrhodotstar}. The redshifts with higher covering fractions are also those that have this transition occur at smaller radii ($z\sim2$ and $z\sim1$ for $k_3=-24.26$). We note that this uncertainty applies to the overall normalization, and is not an uncertainty as a function of redshift.

It is possible to reduce the observed covering fraction by increasing the radius for the transition from atomic to molecular gas. Changing this transition point by $\sim1$kpc for the $z\sim2$ data lowers the observed covering fraction such that it tracks the DLA covering fraction at a $\sim$5\% SFR efficiency. It is therefore possible that the transition point from molecular to atomic-dominated gas may occur at a point further out than we assume in this paper, especially at $z\sim1$ and $z\sim2$. Given this possible systematic, the general agreement of the covering fraction and SFR efficiency shows that there is sufficient DLA gas to account for the emission observed in the outskirts of high redshift SFGs.

\subsection{Covering Fraction of Molecular-dominated Gas}
\label{molecular}

In this section we consider whether the emission observed in the outskirts of SFGs could come from molecular-dominated gas. 
Similar to Section \ref{atomic}, we calculate the covering fraction of molecular-dominated gas using Equation \ref{eq:covfrac}, 
which requires $f(N_{\rm H})$ for molecular-dominated gas ($f(N_{\rm H_2})$). There currently are no measurements of $f(N_{\rm H_2})$ at high redshift, so we use the low redshift measurement and consider possible evolution. Since we do not have direct high-redshift measurements, this section is speculative,
but the results presented in the remainder of the paper are not dependent on the analysis presented in this Section.

\begin{figure*}[t!]
\center{
\includegraphics[scale=0.4, viewport=10 5 490 350,clip]{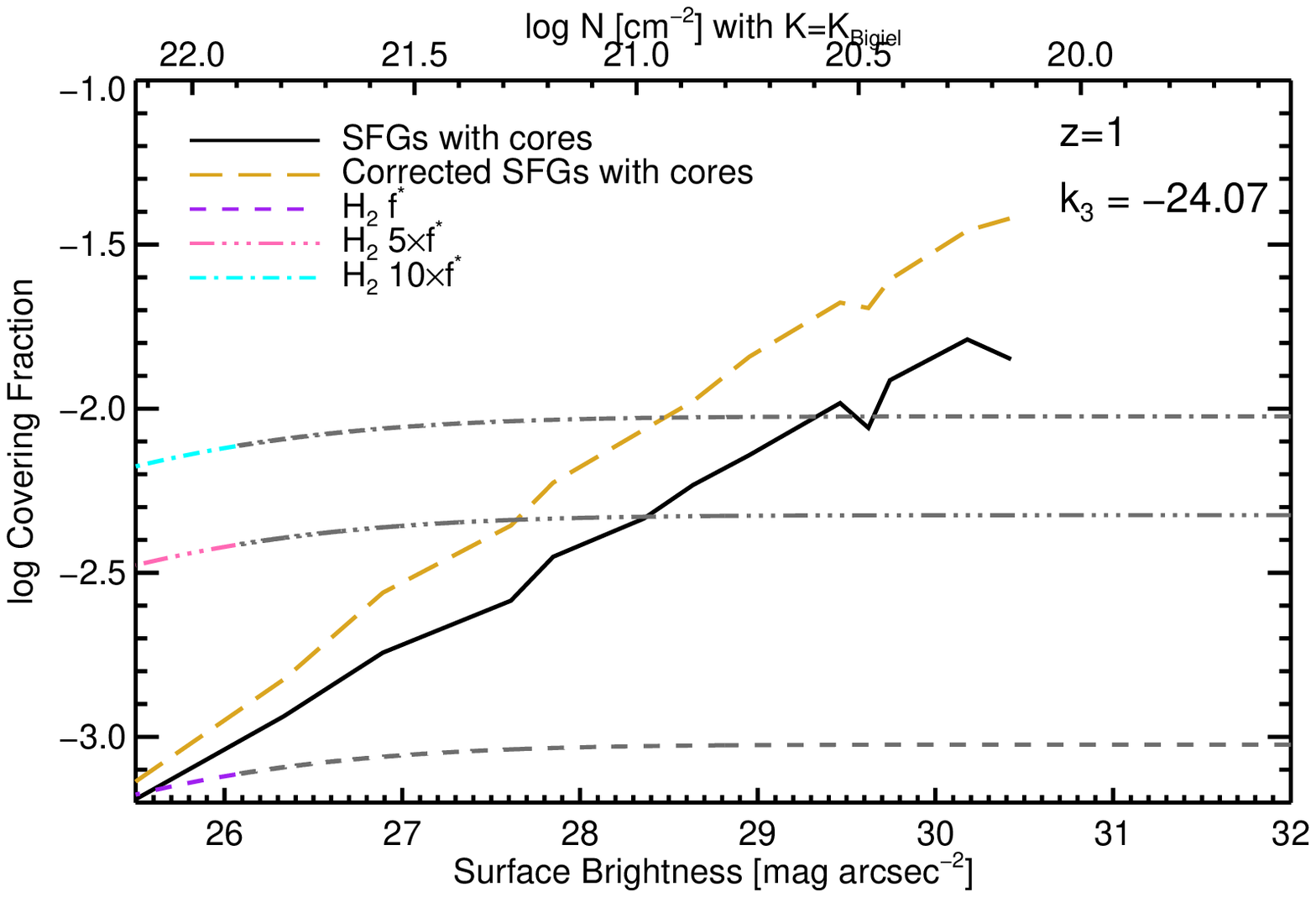}
\includegraphics[scale=0.4, viewport=10 5 490 350,clip]{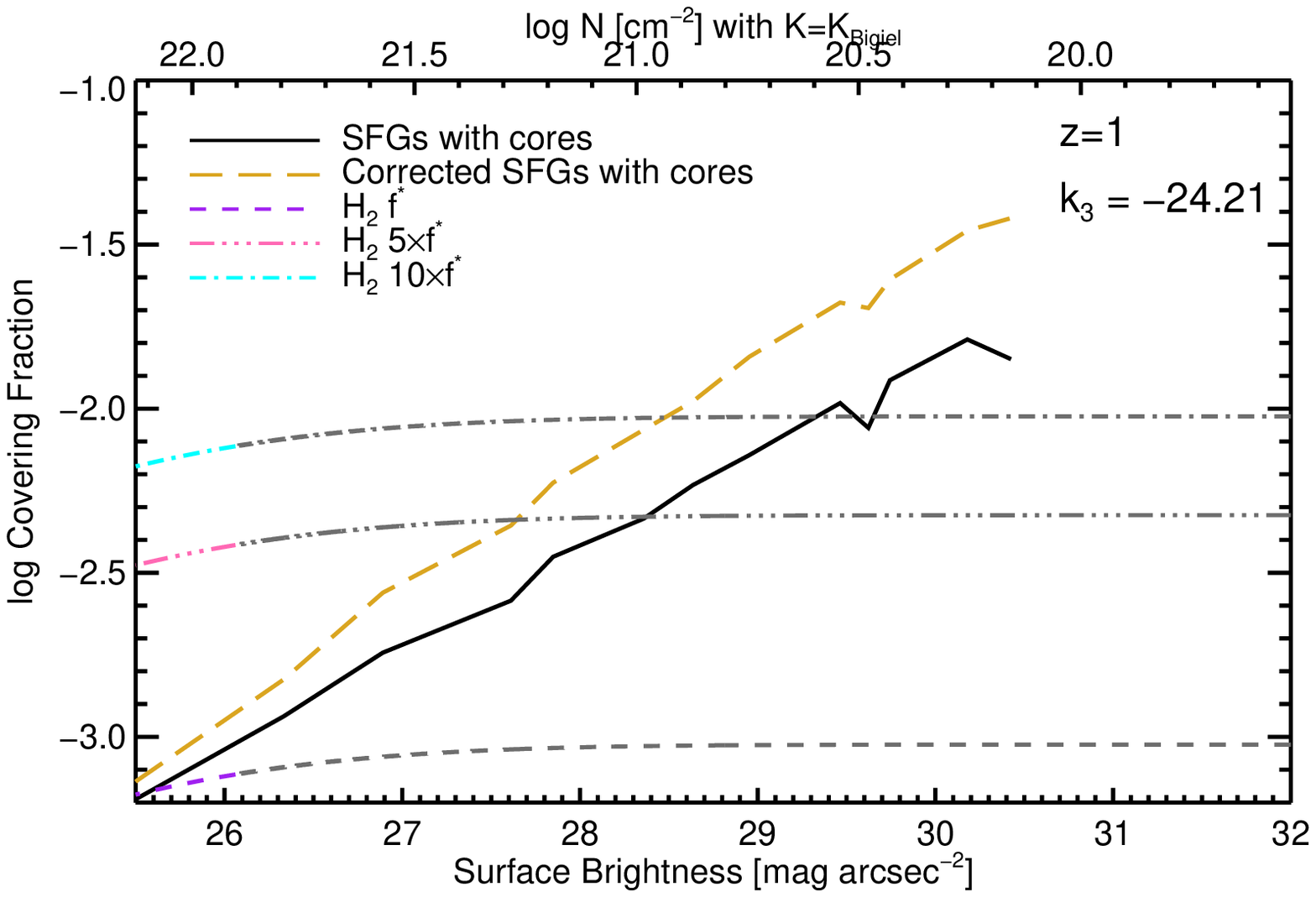}
\includegraphics[scale=0.4, viewport=10 5 490 350,clip]{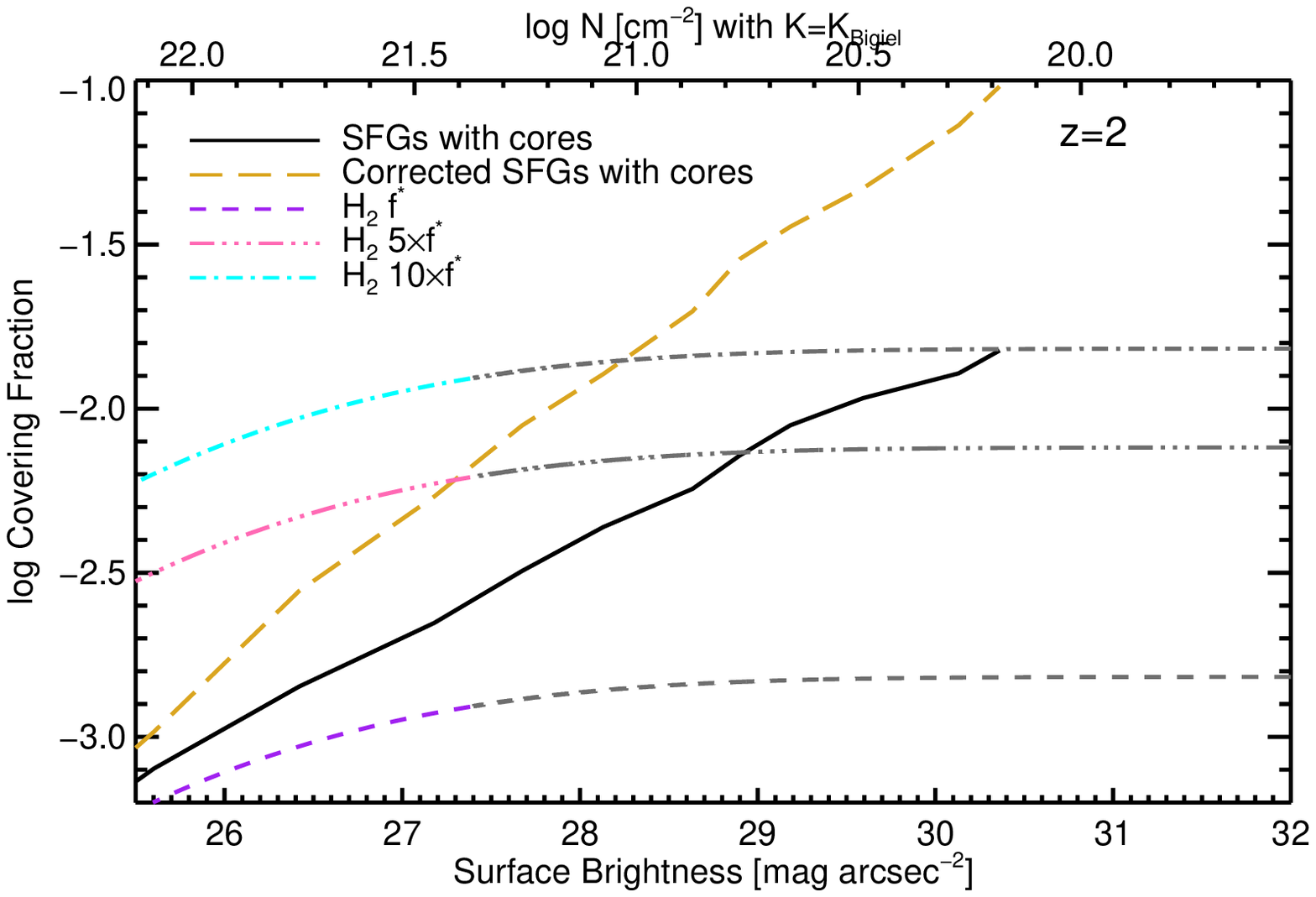}
\includegraphics[scale=0.4, viewport=10 5 490 350,clip]{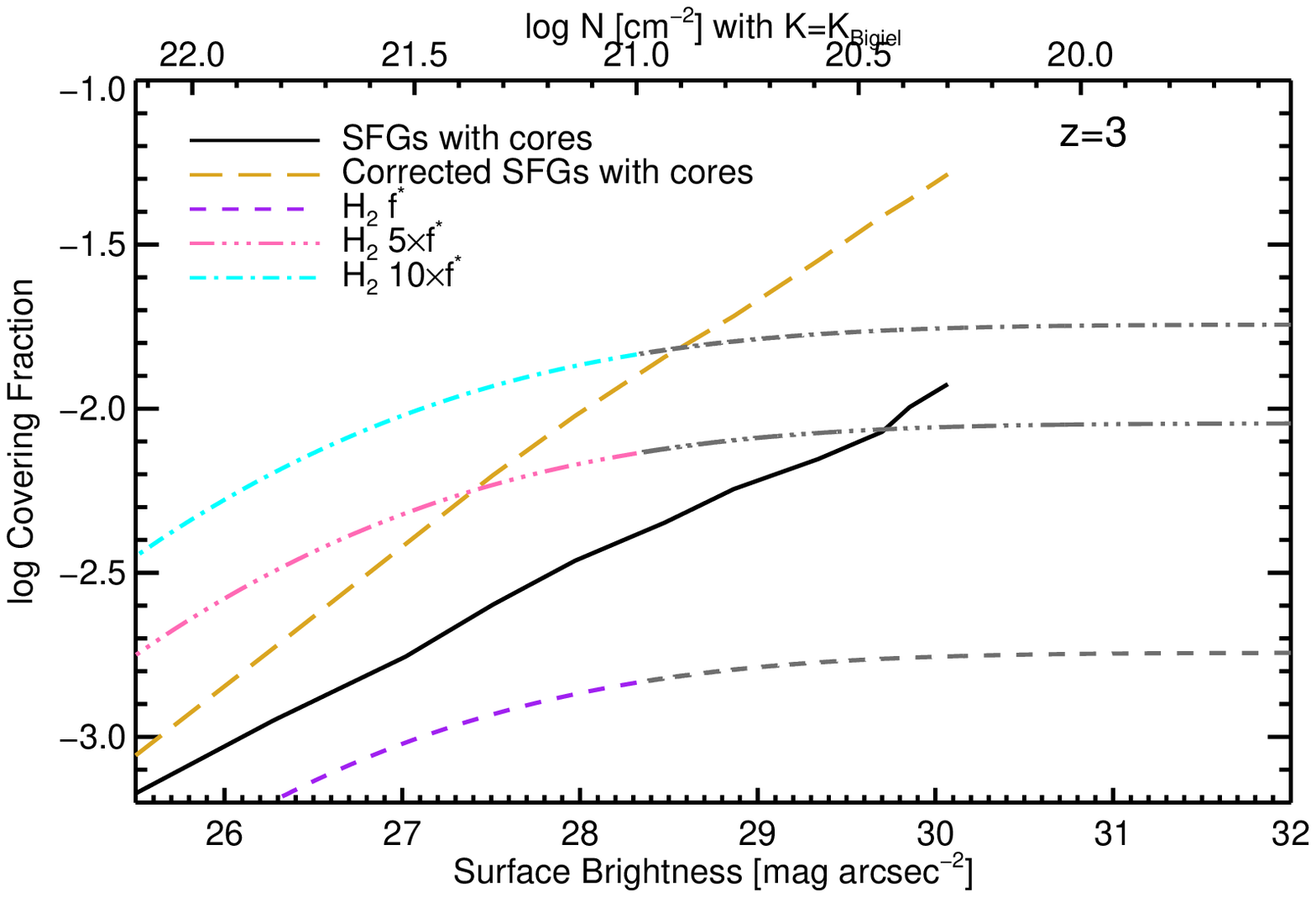}
}
\caption{ \label{fig:covfracH2} 
Cumulative covering fraction of molecular-dominated gas as a function of surface brightness compared to the covering fraction of SFGs. 
This figure is similar to Figure \ref{fig:covfracHI}, but is for molecular-dominated rather than atomic-dominated gas. 
The black line is the covering fraction of the SFGs, and the gold dashed line is the same corrected for completeness.
The purple short-dashed line is the covering fraction of molecular hydrogen with no evolution,
the pink triple-dot-dashed line is the same with an evolution of five times $f^*$ 
(the normalization of the column-density distribution function of molecular gas)
and the cyan dot-dashed line is the same with an evolution of ten times $f^*$.
The gray lines continuing the purple, pink, and cyan lines are extrapolations of the data to lower column densities. 
The column densities labeled at the top of each panel are for $K$=$K_{\rm Biegel}$.
The insufficient covering fraction of molecular-dominated gas at low surface brightnesses suggests that the 
outskirts of SFGs are unlikely to be from molecular-dominated gas. 
}
\end{figure*}

We use the observed $f(N_{\rm H_2})$ from \citet{Zwaan:2006}, who determine a lognormal fit to the BIMA SONG sample \citep{Helfer:2003} defined as:
\begin{equation}
f(N_{\rm H_2})=f^*~{\rm exp}\left[\left(\frac{{\rm log}N-\mu}{\sigma}\right)^2 /~2\right]\;,
\label{eq:lognormal}
\end{equation}

\noindent where $\mu=20.6$, $\sigma=0.65$, and the normalization 
$f^*$ equals $1.1\times10^{-25}$ cm$^2$ \citep{Zwaan:2006}.
We adopt $N_{\rm max} = 10^{24}$, which is the largest observed value of $f(N_{\rm H_2})$.
We use the KS relation with a normalization and slope for molecular gas from \citet{Bigiel:2008},
$K$=$K_{\rm Biegel}$=8.7{$\times$}10$^{-4}$ {\smpykpc} and ${\beta}$=1.0.
From this we can obtain the local covering fraction, which is shown as the purple short-dashed line in Figure \ref{fig:covfracH2}. 
However, given the evolution of the mass density and the SFR density of the Universe, the covering fraction likely evolves between $z=0$ and $z\sim3$, and we consider that possibility below. 

The evolution of $f(N_{\rm H_2})$ would either be due to a change in the slope or normalization, and here we consider
only an evolution in the normalization for the following reasons. First, atomic-dominated gas only appears to evolve in normalization \citep[e.g.][]{Prochaska:2009}.
Second, we have no observational constraints for an evolution in the slope. 
We consider two possible methods to determine the evolution in the normalization. We use the evolution of the mass density 
of molecular gas ($\Omega_{\rm H_2}$) and the SFR density of galaxies. 

The evolution of $\Omega_{\rm H_2}$ is not well constrained by observations, but is carefully explored in models. 
Figure 5 in \citet{Walter:2014} shows the evolution for a range of cosmological models and upper limits from CO observations in 
the Hubble Deep Field North \citep{Obreschkow:2009, Lagos:2011, Lagos:2014, Sargent:2014, Popping:2014}. These results suggest
an evolution of $\Omega_{\rm H_2}$, with an increase from $z=0$ to $z=3$ by a factor of $\sim5$. 

The evolution of the SFR density is better constrained by observations \citep{Schiminovich:2005, Reddy:2009, Bouwens:2015a},
and Figure 18 in \citet{Bouwens:2015a} shows a compilation of the evolution. This suggests an evolution of the SFR density
from $z=0$ to $z\sim2$ of $\sim10$.
The SFR density turns over again at $z\gtrsim2$, 
and hence the maximum evolution is a factor of $\sim10$.

We therefore consider an evolution in the normalization ($f^*$) by a factor of 5 and 10 shown as pink triple-dotted-dashed and cyan dotted-dashed lines in Figure \ref{fig:covfracH2}. 
In addition, there may be molecular-dominated gas down to lower column densities than observed, and thus we extrapolate the data out to lower column densities 
with gray lines continuing each of the three $f(N_{\rm H_2})$ lines. We compare these cumulative covering fractions to that of the SFGs (black lines) in Figure \ref{fig:covfracH2}. The covering fraction 
of the SFGs is different from that shown in Figure \ref{fig:covfracHI}, as it now includes both the outskirts and the inner cores of the galaxies. 
We note that if there is no molecular gas down to these densities, then the covering fraction of molecular gas would be truncated at some surface brightness $<32$ mag in Figure \ref{fig:covfracH2}.

Similar to Section \ref{atomic}, a completeness correction is required, and we use the same procedure described there, except in this case we also include the inner cores of the SFGs in addition to the outskirts. 
The completeness corrected molecular-dominated covering fraction is shown as the gold dashed line in Figure \ref{fig:covfracH2}. This completeness correction is significantly larger than for the atomic-dominated case shown
in Figure \ref{fig:covfracHI}, because the cores of the SFGs are significantly brighter than the outskirts, and therefore the same faint galaxies contribute more at a given surface brightness. 
Since there are a significantly larger number of faint SFGs, and since their cores contribute to the covering fraction for a given surface brightness, this results in a larger overall covering fraction.

Comparing the covering fraction of SFGs to molecular-dominated gas in Figure \ref{fig:covfracH2} shows that there is insufficient molecular-dominated gas for surface brightnesses $\gtrsim28.5$ mag arcsec$^{-2}$, 
assuming a factor of ten increase in $f^*$ based on the evolution of the SFR density. If we instead use the factor of five evolution based on the evolution of $\Omega_{\rm H_2}$, then the covering fraction is insufficient
for surface brightnesses $\gtrsim27.5$ mag arcsec$^{-2}$. We emphasize that a factor of ten increase in $f^*$ is an upper limit to the evolution of $f^*$, as it assumes that all the evolution
in the SFR density is due to an evolution in $f(N_{\rm H_2})$. 

We note that the molecular gas is traced indirectly via CO emission, yet the conversion factor between CO emission and H$_2$, $X_{CO}$, is metallicity dependent, increasing with decreasing metallicity \citep{Wolfire:2010, Bolatto:2011, Bolatto:2013, Elmegreen:2013cn, Amorin:2016}. In addition, there may be a large component of `dark' molecular gas which is missed by tracing H$_2$ with CO observations. Specifically, in low metallicity environments H$_2$ may be shielded, while the CO could be photo-dissociated \citep{Wolfire:2010}. The fraction of this CO-dark gas is also metallicity dependent \citep{Leroy:2011}, and it is not clear how much of the CO-dark gas exists at low metallicity \citep{Langer:2014}. The result is that the molecular content of metal poor gas is poorly constrained, adding significant uncertainty to the covering fraction of molecular gas, which is likely underestimated. As a consequence, the discussion presented in this section is speculative.

Combining the covering fraction results from atomic-dominated gas with those from the molecular-dominated gas shows that atomic-dominated gas can account for the observed emission at $\gtrsim28-29$ mag arcsec$^{-2}$, while
molecular-dominated gas does so at $\lesssim27.5-28.5$ mag arcsec$^{-2}$. This suggests that there may be some overlap in the two, with a possible transition region where star formation occurs in both atomic-dominated and molecular dominated gas. Regardless, it is unlikely that the outskirts of SFGs are forming stars out of molecular-dominated gas since not enough of the sky is covered by this gas to account for the observed emission. 
We note that while this is the case averaged over the H\textsc{i} disks, there could be embedded molecular regions within the disks. In fact, molecular hydrogen is sometimes observed in absorption systems \citep{Noterdaeme:2015, Noterdaeme:2015b}, suggesting this is indeed the case for at least some systems. Further observations of $f(N_{\rm H_2})$ at higher redshift and down to lower gas densities are needed to further this comparison, but it is not a critical part of this analysis. 

Observations of molecular hydrogen in DLAs are very rare, with less than 6\% of the DLA population showing the Lyman and Werner transitions \citep{Jorgenson:2014}. However, when considering a somewhat biased sample of DLAs, there appears to be an increase in the fraction of DLAs with molecular hydrogen as a function of H{\small I} column density \citep{Noterdaeme:2015b}. This is again consistent with the above picture of a transition region with star formation in both atomic-dominated and molecular-dominated gas, as this overlap region would occur at the highest H{\small I} column densities for DLAs. 
  
\section{Discussion}
\label{discussion}

Under the assumption that outskirts of SFGs are composed of atomic-dominated H\textsc{i} gas, the SFR efficiency of this gas is significantly reduced compared to the KS relation, and shows little to no evolution with redshift (see Figure \ref{fig:efficiency}). This assumption is supported by the fact that the covering fraction of molecular-dominated gas is insufficient to explain the emission in the outskirts of SFGs, while the covering fraction of atomic-dominated gas roughly covers the emission. The reduced efficiency is similar to the result by \citet{Rafelski:2011}, and we compare the two studies in Section \ref{comp}. We consider how the normalization $k_3$ of $f(N_{{\rm H  I}},X)$ affects the SFR efficiency in Section \ref{k3}. We interpret the results and compare the results to predictions from models in Section \ref{models}. We also compare them to the SFR efficiency of local H\textsc{i} gas in Section \ref{local}, and to measurements from the double DLA technique in Section \ref{double}. Lastly, we consider the effects of dust in Section \ref{dust}.

\subsection{Comparison to \citet{Rafelski:2011}}
\label{comp}

We compare the SFR efficiency determined here at $z\sim3$ to that previously measured by \citet{Rafelski:2011}, and find that the SFR efficiencies are very similar, although slightly lower than before. Specifically, \citet{Rafelski:2011} find a mean efficiency of $\sim$3.2\%, while the new measurement is 2.8\%. This is well within the $\sim$1\% spread in measurements which vary weakly as a function of the inferred gas surface density. The agreement is not surprising, given that the same technique was implemented in a somewhat overlapping dataset. The key differences include the improved galaxy redshifts from \citet{Rafelski:2015}, the use of both the rest-frame FUV and NUV in the galaxy stacks, an updated $f(N_{{\rm H  I}})$, and the higher sample completeness at fainter magnitudes. 

Even though the improved galaxy redshifts found two catastrophic redshift errors in the \citet{Rafelski:2011} sample, they did not significantly affect the stack since they are created from a median image.
The update to $f(N_{{\rm H  I}})$ at $z\sim3$ was sufficiently small to be inconsequential. Lastly, the higher sample completeness results in less of a dependence on completeness corrections, but the previous corrections were sufficiently accurate that the updated values have not changed significantly. 

\subsection{Normalization of the column density distribution function}
\label{k3}

One of the largest uncertainties in the evolution of the SFR efficiency is the normalization $k_3$ of $f(N_{{\rm H  I}})$ at $z\sim1$. Since $k_3$ is currently not well measured at $z\sim1$, we have to rely on linear fits with redshift as done in Section \ref{fofn} to obtain two possible estimates of $k_3$ at $z=1$. We note that these values of $k_3$ bracket the measurement at $z\sim0.6$.
Figure \ref{fig:efficiency} and Table \ref{tab:SFR} show different SFR efficiencies at $z\sim1$ depending on $k_3$, and we consider those differences here.

One method to obtain $k_3$ at $z\sim1$ is to extrapolate between the $z=0$ and $z\sim2$ measurements, as done with the red dotted line in Figure \ref{fig:fofn_evol}. However, this relies heavily on the $z=0$ value of $k_3$, and this value is highly uncertain as it depends on the shape of $f(N_{{\rm H  I}})$ at $z=0$, which is not well known. Specifically, two studies that measure it report highly discrepant shapes, especially at lower column densities \citep{Zwaan:2005,Braun:2012}.  Moreover, neither one of these shapes agree well with that at $z>2$, although the \citet{Zwaan:2005} measurement is more similar than that by \citet{Braun:2012}. Also, we note that the \citet{Braun:2012} measurements only use a few galaxies to constrain $f(N_{{\rm H  I}})$.

The difference in the shape of $f(N_{{\rm H  I}})$ in the two $z=0$ studies is partially due to differences in the methodology used. It is unclear which method is more correct, and this translates into large systematic uncertainties in $k_3$ at $z=0$. The measurement of $f(N_{{\rm H  I}})$ at $z\sim0.6$ is consistent with the $z>2$ measurements and the \citet{Zwaan:2005} measurement, but not with the \citet{Braun:2012} measurement at lower column densities \citep{Neeleman:2016a}.

We therefore measure $k_3$ at $z=0$ by using the \citet{Zwaan:2005} data to fit $f(N_{{\rm H  I}})$, forcing the shape to match that at $z>2$. This is not a good fit due to the shape disagreement, but it provides a reasonable estimate for $k_3$ of $-24.07$, which corresponds to the red crosses in Figure \ref{fig:efficiency}. For this value of $k_3$ there is a slight decrease in the SFR efficiency of H\textsc{i} gas with decreasing redshift.
We note that this is not possible with the \citet{Braun:2012} data, as the shape is too deviant from that at $z>2$. 

An alternative (and preferred) method is to obtain $k_3$ at $z\sim1$ by a linear fit of the data including the $z=0.6$ and $z>2$ data together, as shown with the blue dashed line in Figure \ref{fig:fofn_evol}. This results in $k_3=-24.21$, which corresponds to the blue diamonds in Figure \ref{fig:efficiency}. This result is also consistent with the measurement at $z\sim0.6$ \citep{Neeleman:2016a}, although the higher redshift data points are more heavily weighted due to the large uncertainty. An extrapolation of fitting just the $z>2$ data yields $k_3=-24.26$, which results in a very similar SFR efficiency and thus is not shown (but is tabulated in Table \ref{tab:SFR}). 
This method of obtaining $k_3$ is likely more reliable than that obtained from the extrapolation to the highly uncertain measurement at $z=0$, as it is based on measurements conducted in the same fashion. For this measurement of $k_3$, there is no observed evolution in the SFR efficiency of H\textsc{i} gas with redshift, and we consider it as our primary measurement of the SFR efficiency below.  

\subsection{Comparison with models}
\label{models}

Galaxy simulations incorporating H$_2$-regulated star formation find that the primary source for lower SFR efficiencies at high redshift is a decrease in the dust-content, which is well traced by the metallicity of the gas \citep{Gnedin:2010, Kuhlen:2012, Hopkins:2014dn,Somerville:2015b}. These simulations predict a reduced SFR efficiency for DLA gas due to their low metallicities. For instance, \citet{Gnedin:2010} predict efficiencies matching those by \citet{Rafelski:2011} at $z\sim3$. The low dust content of DLAs, as traced by the metallicity, could therefore explain the reduced SFR efficiencies.  In fact, studies of star formation at low metallicity in the Small Magellanic Cloud also find a lower SFR efficiency \citep{Bolatto:2011, Jameson:2015}.

\begin{figure}[]
\center{
\includegraphics[scale=0.5, viewport=10 5 490 355,clip]{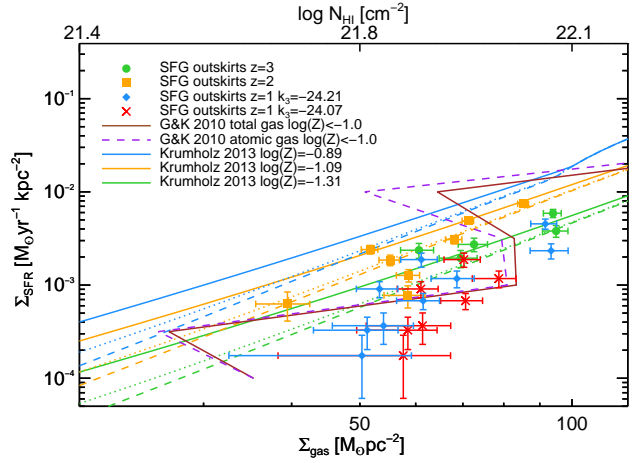}
}
\caption{ \label{fig:krum} 
Comparison of the SFR efficiency of DLA gas in the outskirts of SFGs at $z\sim1$-3 with the cosmological simulation at $z\sim3$ by \citet{Gnedin:2010} and the analytic KMT+ model by \citet{Krumholz:2013a} for molecule-poor galaxies for different metallicities.
The brown line is the mean relation for the total neutral-gas surface density (atomic and molecular) in the cosmological simulation, and the purple dashed line is the same for only atomic hydrogen gas.
The KMT+ model predictions are for the metallicities of DLAs at the redshifts of the binned SFGs, with the colors matching those of the corresponding measurements at each redshift:
green for $z=3$, orange for $z=2$, and blue for $z=1$. The solid, dotted, and dashed lines use stellar densities of 0.1 M$_\odot$ pc$^{-3}$, 0.01 M$_\odot$ pc$^{-3}$, and 0.001 M$_\odot$ pc$^{-3}$ respectively. 
The model provides a measure of the expected evolution in the SFR as a function of redshift due to the metallicity evolution of DLAs \citep{Rafelski:2012},
which would evolve from the green line at $z=3$ to the blue line at $z=1$. 
Our measurements use the same symbols as in Figure \ref{fig:efficiency}, consisting of green circles, orange squares, blue diamonds, and red crosses respectively. The data are broadly in agreement with the simulations and models, although the data do not reproduce the predicted metallicity evolution. Specifically, the blue diamonds and red crosses are predicted to fall on the blue line, and they fall significantly below and to the right, revealing no evolution with redshift and therefore metallicity. 
}
\end{figure}

We compare our results to the cosmological simulations by \citet{Gnedin:2010} at $z\sim3$ for gas with metallicity below 0.1 $Z_\odot$\footnote{The metallicity cut of 0.1 $Z_\odot$ is reasonable, given the mass-weighted and volume-weighted metallicity of atomic gas is $\sim0.02$ $Z_\odot$ and $\sim0.03$ $Z_\odot$ respectively \citep{Rafelski:2011}.}. These simulations include a metallicity dependent model of molecular hydrogen \citep{Gnedin:2009}. Figure \ref{fig:krum} shows that the \citet{Gnedin:2010} simulations predict the same SFR efficiency as measured here. \citet{Gnedin:2010} conclude that the lower metallicity, and therefore lower dust-to-gas ratio, causes a decrease in the amplitude of the KS relation, as observed here.  \citet{Gnedin:2010} also find that while the higher UV flux at high redshift does lower the SFR, it also lowers the surface density of the neutral gas, leaving the KS relation mostly unaffected. 

Due to the limited redshift and metallicity range of these simulations, we can not directly investigate the expected evolution as a function of redshift and metallicity. We therefore turn to analytic models. The KMT+ model by \citet{Krumholz:2013a} explicitly computes the behavior of H\textsc{i}-dominated gas in the galaxy outskirts, and provides us with the $\Sigma_{\rm SFR}$ as a function of the metallicity and $\Sigma_{\rm gas}$. This model is based on the KMT slab model \citep{Krumholz:2008, Krumholz:2009a,Krumholz:2009b}, which resulted in a fixed ratio of the interstellar radiation field (ISRF) to gas density. 
The newer KMT+ model allows the ISRF to gas density ratio to vary, which is necessary at low ISRF intensities to maintain hydrostatic equilibrium. This results in a floor on the density of cold  H\textsc{i} gas, and also results in a floor on the H$_2$ fraction and the SFR \citep{Krumholz:2013a}.

The analytic model by \citet{Krumholz:2013a} is able to reproduce the observed low efficiencies at the metallicities of DLAs.
Moreover, the model can also be used to predict the expected $\Sigma_{\rm SFR}$ as a function of the  $\Sigma_{\rm gas}$ at different metallicities, and therefore redshifts. The metallicity of DLA gas increases with decreasing redshift, with a $\sim0.4$ dex evolution from $z\sim3$ to $z\sim1$ \citep[log($Z$)$=-1.31$ to log($Z$)$=-0.89$, where $Z$ is the ratio of the metal to hydrogen column density normalized to solar;][]{Rafelski:2012, Rafelski:2014}. This translates to a change of $\sim0.5$ dex in the $\Sigma_{\rm SFR}$, assuming constant stellar densities \citep{Krumholz:2013a}, as shown in Figure \ref{fig:krum}. While this level of evolution can be measured with the precision of our measurements, we find no evolution in the SFR efficiency of atomic-dominated H\textsc{i} gas. While the model is somewhat consistent with the evolution observed from $z\sim3$ to $z\sim2$, the $z\sim1$ points fall significantly below and to the right of the model. This is true regardless of the value of $k_3$ used in $f(N_{{\rm H  I}})$ at $z\sim1$.

Besides gas densities and metallicity, stellar density is an additional parameter that in some models \citep[e.g.][]{McKee:2007} regulates the SFR, although they are unknown for DLAs.
To gauge the importance of stellar density, we turn again to the model by \citet{Krumholz:2013a}, computing the expected SFR for three values of the stellar density, 0.1 M$_\odot$ pc$^{-3}$, 0.01 M$_\odot$ pc$^{-3}$, and 0.001 M$_\odot$ pc$^{-3}$. The local neighborhood has stellar densities of 0.01 M$_\odot$ pc$^{-3}$ \citep{Holmberg:2000}, and increasing or decreasing the stellar density does not alter our conclusions substantially as shown in Figure \ref{fig:krum}. In addition, the gas surface densities are sufficiently high such that the stellar gravity is a small contribution to the total pressure, and therefore they should have small effects on the resultant SFR efficiencies \citep{Krumholz:2013a}. These surface densities are also unlikely to evolve significantly from $z\sim2$ to $z\sim1$. 

In the KMT+  model, the FUV background is dominated by the local star formation rather than the cosmic FUV background \citep{Krumholz:2013a,Haardt:2012}, and the FUV field and SFR are computed self-consistently. Therefore any feedback effects caused by the decreased UV radiation from the lower SFR efficiency as a function of metallicity is already built into the model. However, this depends on whether we truly understand the interplay between the metallicity and the UV background, and how the two play off each other in cosmological simulations \citep[e.g.][]{Gnedin:2010, Gnedin:2011, Somerville:2015b}. For instance, if the FUV radiation field is not dominated by local star formation, but rather by cosmic FUV background, then it may be possible that the lack of evolution could be caused by a balance of an increased efficiency of higher metallicity gas cancelled by the lower-efficiency of a higher cosmic FUV background. However, this is not currently predicted by models. 

There are many uncertainties in how we implement the UV background in models and simulations, which would affect the amplitude of the UV background. Moreover, we do not fully understand the interplay between the metallicity and the UV background, and how the two play off each other in a cosmological context (Rachel Somerville, private communication 2016). Further investigations with cosmological simulations is warranted, and such simulations would need to reproduce our observed lack of evolution in the SFR efficiency with redshift and metallicity from $z\sim1-3$. 

One possibility that could affect the SFR efficiency comparison is if we had to exclude from our analysis a population of ``low cool'' DLAs \citep{Wolfe:2008}, i.e. a subset of DLAs which are believed to be unrelated to ongoing star formation. Excluding low-cool DLAs would decrease the number of DLAs (or equivalently reduce the value for $f(N_{{\rm H  I}},X)$), yielding a lower $d\dot{\rho_*}/{d\langle I_{\nu_0}^{\rm obs}\rangle}$. In \citet{Rafelski:2011} we find that this decrease would result in a $\sim0.1$ dex shift to the left in the inferred $\Sigma_{\rm gas}$ in Figure \ref{fig:krum}, which also brings the measurements into somewhat better agreement with the model. While no redshift evolution of this bimodality is likely \citep{Wolfe:2008}, even such an evolution would not be sufficient to change our result to an evolution in the SFR efficiency with redshift.

\subsection{Comparison with local data}
\label{local}

An alternative explanation for the reduced SFR efficiency is the role of molecular versus atomic hydrogen gas in star formation. Empirically, the $\Sigma_{\rm SFR}$ of local spiral galaxies is well correlated with the $\Sigma_{\rm H_2}$ \citep[e.g.][]{Bigiel:2008, Bigiel:2011}, and some have argued that the KS-relation is only valid for molecular-dominated gas rather than for atomic-dominated gas \citep[e.g.][]{Wong:2002, Schruba:2011}. However, an analysis of the atomic-dominated gas in the outskirts of spiral galaxies reveals a clear correlation of the $\Sigma_{\rm SFR}$  and the  $\Sigma_{\rm HI}$ \citep{Bigiel:2010b, Bigiel:2010a, Roychowdhury:2015}. Similarly, atomic-dominated dwarf galaxies also have a clear correlation of the $\Sigma_{\rm SFR}$ with the  $\Sigma_{\rm HI}$ \citep{Roychowdhury:2014, Elmegreen:2015}. 

We do not assert that stars form directly out of atomic-dominated H\textsc{i} gas, and our measurements do not distinguish nor depend on whether the atomic hydrogen gas transitions to the molecular phase before forming stars. However, some simulations show that gas without H$_2$ or CO can cool to low enough temperatures to form stars by gravitational collapse \citep{Glover:2011, Glover:2012a}. The H$_2$ observed at low metallicity may therefore be a consequence of star formation, rather than the cause. It is very difficult to measure H$_2$ in these regimes, since much of this H$_2$ may be CO-dark \citep{Wolfire:2010}, and the covering fraction of H$_2$ may be lower than that of H\textsc{i}. Even so, we expect that there will be some molecular gas present in these atomic-dominated regimes \citep{Schruba:2011}, and it is even observed directly in one low-metallicity dwarf galaxy \citep{Leroy:2006cy}. 

DLAs have a sufficiently high column density of H\textsc{i} gas that the gas is self-shielded \citep{Wolfe:2005}, which could allow star formation even with low levels of dust. 
The main constraint for DLA gas is that molecular hydrogen has a very small cross-section either due to a lack of molecular hydrogen in the gas, the molecular phase is present over only a short time scale, or the volume filling fraction is small, since most DLAs do not show evidence of abundant H$_2$ based on the Lyman and Werner transitions \citep{Jorgenson:2014,Noterdaeme:2015b}.

\begin{figure}[]
\center{
\includegraphics[scale=0.5, viewport=10 5 490 355,clip]{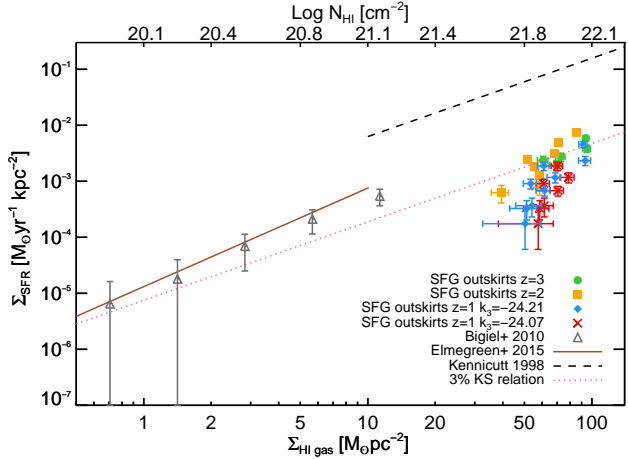}
}
\caption{ \label{fig:local} 
Comparison of the SFR efficiency of DLA gas in the outskirts of SFGs at $z\sim1$-3  with local measurements. 
This shows that the local spiral galaxy outskirts and dwarf galaxies have a similar reduced SFR efficiency as measured here statistically in DLAs.
The pink dotted line represents the KS relation at 3\% efficiency, which is consistent with our measurements and similar to the local measurements, although slightly lower.
Our measurements again use the same symbols as in Figure \ref{fig:efficiency}, consisting of green circles, orange squares, blue diamonds, and red crosses. 
The gray triangles are for the outskirts of local spiral galaxies \citep{Bigiel:2010b}, the brown line is for local dwarf galaxies \citep{Elmegreen:2015}, and the black line
represents the KS relation \citep{Kennicutt:1998a}.  }
\end{figure}

Figure \ref{fig:local} compares the SFR efficiency measured in DLA gas in the outskirts of SFGs to that of the H\textsc{i} gas in the outskirts of local spiral \citep[][gray triangles]{Bigiel:2010b} and dwarf galaxies \citep[][brown line]{Elmegreen:2015}, and shows a reduced SFR efficiency in atomic-dominated H\textsc{i} gas compared to the KS relation, shown as the black dashed line. These measurements at $z=0$ show SFR efficiencies similar (slightly higher) to those presented here at $z\sim$1-3 (e.g. pink dotted line in Figure \ref{fig:local}), suggesting that perhaps the efficiency of this gas is not changing significantly over time or strongly with metallicity. 

The similar efficiency of H\textsc{i}-dominated gas at both low and high redshift and at different metallicities shown in Figure \ref{fig:local}, combined with the lack of any observed evolution in the efficiency of DLAs as a function of redshift (and therefore metallicity), suggests that the reduced SFR efficiency is likely driven by the low molecular content of the atomic-dominated phase. At the same time, the metallicity could play a secondary effect on the efficiency by regulating the conversion between atomic and molecular gas. Theory predicts that there is a metallicity dependence on this transition \citep{Krumholz:2009a}, which would explain the observed high column-densities of DLAs without much molecular gas.
In this scenario, the metallicity of the DLA gas is too low for the atomic gas to form significant amounts of molecular gas at the given surface densities. This may explain how the KMT+ model does not reproduce the lack of evolution in the efficiency of the DLA gas, while at the same time reproducing  the SFR efficiency of the local H\textsc{i} gas \citep{Krumholz:2013a}. We note that our data do not directly constrain the metallicity dependence of the transition to the molecular phase. Rather, we add constraints for future models and simulations by providing measurements that show no strong evolution in the SFR efficiency as a function of metallicity. 

\vspace{5mm}
\subsection{Comparison with double DLA measurements}
\label{double}

The dependence of the  H\textsc{i} to H$_2$ transition on metallicity coupled with the typically higher metallicity at $z=0$ suggests that it is unlikely that we will measure H\textsc{i} gas locally at the surface densities of the high column density DLA gas measured here.
However, it is possible to measure the SFRs of DLAs at lower column densities at high redshift using the double DLA technique described in Section \ref{intro}.
This technique was used to search for emission from typical DLAs at $z\sim2-3$ \citep[][]{Fumagalli:2010a, Fumagalli:2014b, Fumagalli:2015a}, 
and no emission from the DLAs was detected.

\begin{figure}[]
\center{
\includegraphics[scale=0.5, viewport=10 5 490 355,clip]{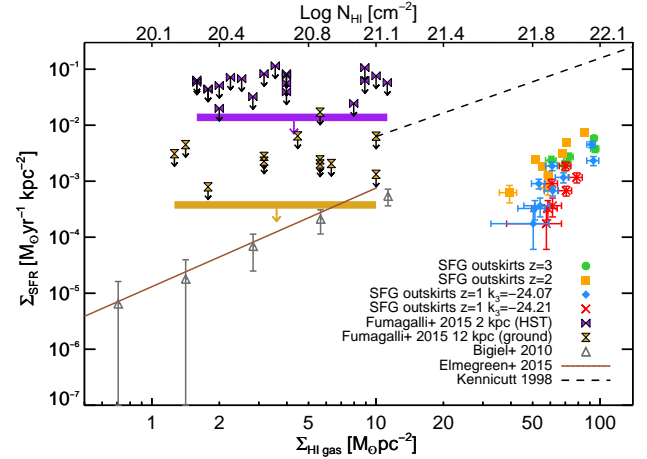}
}
\caption{ \label{fig:double} 
Comparison of the SFR efficiency of DLA gas in the outskirts of SFGs at $z\sim1$-3  with local measurements and 
lower column-density DLA gas via the double DLA technique. 
This shows that the upper limits from direct DLA 
observations are consistent with both our measurements and the local H\textsc{i} gas.
The upper limits of DLAs at $z\sim2-3$ for HST and ground observations are shown in purple and gold bowties respectively \citep{Fumagalli:2015a}.
Our measurements use the same symbols as in Figure \ref{fig:efficiency}, consisting of green circles, orange squares, blue diamonds, and red crosses. 
The gray triangles are for the outskirts of local spiral galaxies \citep{Bigiel:2010b}, the brown line is for local dwarf galaxies \citep{Elmegreen:2015}, and the black line
represents the KS relation \citep{Kennicutt:1998a}. }
\end{figure}

The upper limits of these measurements for HST and ground observations are shown in purple
and gold bowties in Figure \ref{fig:double}. In addition, these measurements were stacked to improve signal-to-noise, shown
as the solid purple and gold horizontal bars. The HST and ground measurements cannot be combined, as they measure star
formation over different areas matched to the resolution of the images. We note that these measurements determine $\Sigma_{\rm SFR}$ over a specific area,
and these values could be higher if star formation is occurring only in regions smaller than these apertures, which are matched to the resolution of the telescopes. 
Also, we note that these limits are not corrected for inclination. 

The resulting upper limits are consistent both with the measurements presented here,
and those of local spiral and dwarf galaxies. However, the sensitivities are not sufficient to yield strong constraints on the SFR efficiency of the gas. Also, we note that the results and model from this study are consistent with the results in \citet{Fumagalli:2015a}, as the galaxies studied here are for the most part below their sensitivity level. 
Measurements at higher H\textsc{i} column densities with higher sensitivities would enable a direct comparison
with the results from this study, and are likely to result in direct detections of typical DLAs. This would in turn result in a more direct measurement of the SFR efficiency of atomic-dominated H\textsc{i} gas at high redshift.

\subsection{Possible dust extinction}
\label{dust}

The rest-frame FUV light from normal high-redshift SFGs suffers from dust extinction by a factor of up to five \citep{Reddy:2012a}, which could reduce our measured SFR efficiency. However, in the outskirts of local galaxies, \citet{Bigiel:2010b} find that the FUV emission reflects the recently formed stars without large biases from external extinction. We similarly do not expect much extinction in the outer parts of the SFGs, as DLAs have low dust-to-gas ratios and low levels of extinction \citep{Murphy:2004, Frank:2010, Khare:2012, Fukugita:2015, Murphy:2016}. 

Regardless, if we consider the unlikely scenario in which the maximum extinction were present in the outskirts, the SFR efficiency would still not be close to the expected value, and the reduced SFR efficiencies would remain. Specifically, a factor of five in extinction would raise the SFR efficiencies by approximately a factor of five as well, resulting in efficiencies of $\lesssim20$\%. However, this is unlikely since the dust is concentrated in the center of the galaxies \citep{Nelson:2016}.

Another unlikely possibility to consider is if dust extinction could cause us to miss an evolution with redshift. This would require that the outskirts of $z\sim1$ SFGs have more dust than the $z\sim2$ or $z\sim3$ SFGs. To test this possibility, we artificially added extinction to the $z\sim1$ stack, and find that a minimum of $\sim1$ mag extra extinction over that at $z\sim2$ would be necessary to make the measurements consistent with a metallicity evolution, and an extinction of $\sim3$ mags would be needed to clearly show an evolution with redshift. This assumes that log $k_3=-24.26$, while log $k_3=-24.07$ would require even more extinction. 

It is extremely unlikely that the extinction in the outskirts of these galaxies would evolve so drastically from $z\sim2$ to $z\sim1$, especially given the low levels of extinction measured in DLAs at similar redshifts. While there is a small spread in the reddening caused by DLAs, the most recent measurements find E(B-V) = 0.003 mag \citep{Murphy:2016}, while the highest measurement is E(B-V) = 0.01 mag \citep{Fukugita:2015}. We note that while the extinction is correlated with the metallicity of DLAs, it is uncorrelated with $N_{\rm HI}$ \citep{Murphy:2016}, which is consistent with the finding that $N_{\rm HI}$ in DLAs is uncorrelated with the metallicity of DLAs \citep{Neeleman:2013}. We therefore conclude that dust extinction does not affect our results.

\section{Summary}
\label{summary}
 
Measurements of the evolution in the SFR efficiency of neutral atomic-dominated hydrogen gas are needed to understand 
the origin of the reduced SFR efficiency measured at $z\sim3$ \citep{Wolfe:2006, Rafelski:2011}. 
Here we present the SFR efficiency of this gas at $z\sim1$, $z\sim2$, and $z\sim3$ measured in the outskirts of SFGs, assuming that DLAs are associated with SFGs, and that the outskirts of these galaxies are composed of H\textsc{i} gas. These assumptions are well warranted, given the significant evidence for the association of DLAs with SFGs, and the covering fractions of the outskirts of these galaxies compared to the molecular and atomic-dominated gas (Figures \ref{fig:covfracHI} and \ref{fig:covfracH2}). 

We select SFGs in the UDF at $z\sim1$, $z\sim2$, and $z\sim3$ using new photometric redshifts from \citet{Rafelski:2015} utilizing new UV imaging of the UDF, and create composite image stacks of isolated, compact, and symmetric SFGs in the rest-frame UV. We extract radial surface brightness profiles (Figure \ref{fig:sbprofiles}), which show low surface-brightness emission out to large radii ($\sim8$ kpc). This emission is interpreted as in-situ star formation in H\textsc{i} gas surrounding the SFGs. 

In order to obtain the SFR efficiency, we require $f(N_{\rm HI})$, and we fit a double power-law to the number of high density absorbers to background quasars at $z\sim2-3.5$ using the data from \citet{Noterdaeme:2012} combined with the $z=0.6$ data from \citet{Neeleman:2016a} to determine the evolution of the normalization $k_3$ as a function of redshift. In addition, we consider a second value of $k_3$ by interpolating between the $z\sim2$ value, and the $z\sim0$ value obtained by fitting double power-law to the data from \citet{Zwaan:2005}, requiring the same shape. Figure \ref{fig:fofn_evol} shows the evolution for both possible values of $k_3$, and we consider both possibilities for $k_3$ when determining the SFR efficiency. 

We determine the SFR efficiency of atomic-dominated hydrogen gas at each redshift by comparing the emission in the outskirts of SFGs to that expected from the \citet{Rafelski:2011} model, which is based on the KS relation and $f(N_{\rm HI})$ (see Figure \ref{fig:diffrhodotstar}). The resulting SFR efficiencies are similar to those found at $z\sim3$ by \citet{Rafelski:2011}. The SFR efficiency is then combined with the measured $\Sigma_{\rm SFR}$ to visualize the results on the familiar KS relation plot (Figure \ref{fig:sfr_sigma}). We directly compare the SFR efficiencies at different redshifts to each other as a function of their inferred $\Sigma_{\rm gas}$ to investigate any evolution in the efficiency (Figure \ref{fig:efficiency}), and find no evolution with redshift.

Given the success of models and simulations using H$_2$-regulated star formation in reproducing the reduced SFR efficiency of the H\textsc{i} gas at $z\sim3$ \citep{Gnedin:2010, Krumholz:2013a}, the expectation was that the SFR efficiency would evolve with the metallicity. The metallicity of DLA gas increases with decreasing redshift, with a $\sim0.4$ dex evolution from $z\sim1$ to $z\sim3.0$ \citep{Rafelski:2012, Rafelski:2014}, which would be more than sufficient to measure an increase in the SFR efficiency of the gas (Figure \ref{fig:krum}). However, the lack of any evolution of the SFR efficiency with redshift instead suggests that it may be driven by the low molecular content of this atomic-dominated phase. The metallicity may instead play a secondary effect in regulating the conversion between atomic and molecular gas. This interpretation is supported by the similar SFR efficiency observed in H\textsc{i} gas at $z=0$ in the outskirts of spiral galaxies \citep{Bigiel:2010b} and dwarf galaxies \citep{Elmegreen:2015} (Figure \ref{fig:local}). 

We compare the results measured statistically here to direct measurements of individual DLAs using the double DLA technique \citep{Fumagalli:2015a}. We find that the two results are consistent with each other, as the direct measurements are all upper limits at lower H\textsc{i} column densities (Figure \ref{fig:double}). A more direct comparison would be possible with deep UV imaging targeting higher column density gas using the double DLA technique. Such observations are currently possible with WFC3/UVIS on the Hubble Space Telescope, and would verify the results presented here, and yield the first direct imaging of a typical DLA unbiased by metallicity. 

\acknowledgments

 We would like to thank Rachel Somerville, Mark Krumholz, and Bruce Elmegreen 
 for useful discussions on interpreting the results. We also thank the referee for useful comments that improved the clarity of the paper.
 MR acknowledges support from an appointment to the NASA Postdoctoral Program at Goddard Space Flight Center.
 MF acknowledges support by the Science and Technology Facilities Council [grant number ST/L00075X/1].
 Support for HST Program GO-12534 was provided by NASA
 through grants from the Space Telescope Science Institute, which is
 operated by the Association of Universities for Research in
 Astronomy, Inc., under NASA contract NAS5-26555.

 \facility{} Facilities:  HST (WFC/ACS, WFC3/UVIS, WFC3/IR)

\bibliography{sfe}

\end{document}